\documentclass[twocolumn,twocolappendix]{aastex631}
\usepackage{amsmath}

\received{}
\revised{}
\accepted{}

\submitjournal{The Astrophysical Journal}

\shorttitle{HOW-MHD}
\shortauthors{Seo and Ryu}

\begin{document}

\title{HOW-MHD: A High-Order WENO-Based Magnetohydrodynamic Code\\ with a High-Order Constrained Transport Algorithm for Astrophysical Applications}
\author[0000-0002-5550-8667]{Jeongbhin Seo}
\affiliation{Department of Physics, College of Natural Sciences, UNIST, Ulsan 44919, Korea}
\author[0000-0002-5455-2957]{Dongsu Ryu}
\affiliation{Department of Physics, College of Natural Sciences, UNIST, Ulsan 44919, Korea}

\correspondingauthor{Jeongbhin Seo}
\email{jeongbhinseo@unist.ac.kr}
\correspondingauthor{Dongsu Ryu}
\email{dsryu@unist.ac.kr}

\begin{abstract}

Due to the prevalence of magnetic fields in astrophysical environments, magnetohydrodynamic (MHD) simulation has become a basic tool for studying astrophysical fluid dynamics. To further advance the precision of MHD simulations, we have developed a new simulation code that solves ideal adiabatic or isothermal MHD equations with high-order accuracy. The code is based on the finite-difference weighted essentially non-oscillatory (WENO) scheme and the strong stability-preserving Runge-Kutta (SSPRK) method. Most of all, the code implements a newly developed, high-order constrained transport (CT) algorithm for the divergence-free constraint of magnetic fields, completing its high-order competence. In this paper, we present the version in Cartesian coordinates, which includes a fifth-order WENO and a fourth-order five-stage SSPRK, along with extensive tests. With the new CT algorithm, fifth-order accuracy is achieved in convergence tests involving the damping of MHD waves in three-dimensional space. And substantially improved results are obtained in magnetic loop advection and magnetic reconnection tests, indicating a reduction in numerical diffusivity. In addition, the reliability and robustness of the code, along with its high accuracy, are demonstrated through several tests involving shocks and complex flows. Furthermore, tests of turbulent flows reveal the advantages of high-order accuracy, and show the adiabatic and isothermal codes have similar accuracy. With its high-order accuracy, our new code would provide a valuable tool for studying a wide range of astrophysical phenomena that involve MHD processes.

\end{abstract}


\section{Introduction}
\label{s1}

There are many astrophysical objects and phenomena where magnetic fields in conducting fluids play crucial roles.  Examples include accretion disks \citep[e.g.,][]{balbus1998,sorathia2012}, turbulence and star formation in the interstellar medium (ISM) \citep[e.g.,][]{cho2000,elmegreen2004,padoan2011}, astrophysical jets \citep[e.g.,][]{oneill2005,zanni2007}, the intracluster medium (ICM) of galaxy clusters \citep[e.g.,][]{dubois2008,porter2015}, and also solar flares and winds \citep[e.g.,][]{janvier2015,gombosi2018}. To describe them, based on the so-called MHD approximation, magnetohydrodynamic (MHD) equations were derived \citep[see, e.g.,][]{jackson1962,shu1992b} and have been applied.

In applications of MHD equations, the adiabatic equation of state (EoS) is commonly employed, where the pressure $P$ changes with the fluid density $\rho$ as $P\propto\rho^{\gamma}$, away from shocks. Here, $\gamma$ is the adiabatic index. This EoS holds if cooling is negligible or the cooling timescale is much longer than the dynamical timescale. On the other hand, in the limit where the cooling timescale is much shorter than the dynamical timescale, the assumption of isothermal flows, in which the fluid temperature $T\propto P/\rho$ is set to be constant, becomes physically plausible and may be used \citep[see, e.g.,][]{draine1993}. This isothermal EoS is widely used in studies of turbulent flows \citep[e.g.,][]{ostriker2001,kritsuk2007,federrath2008,porter2015,roh2019}. In cases where the cooling timescale is comparable to the dynamical timescale, cooling must be considered explicitly.

Due to the inherent complexity of MHDs, numerical simulations have become the standard method for studying MHD processes. A large number of simulation codes for MHD equations have been developed, and are used in different fields of research. A partial list of publicly available MHD codes, applied primarily to astrophysical problems, includes ZEUS based on an scheme using artificial viscosity \citep{stone1992a}, FLASH based on the eight-wave model \citep{powell1999} or an unsplit staggered mesh algorithm \citep{lee2009}, PLUTO based on a Godunov-type scheme with the HLLD Riemann solver \citep{mignone2007}, ATHENA based on a higher-order Godunov method \citep{stone2008}, and CANS+ based on a fifth-order-monotonicity-preserving interpolation (MP5) scheme with the HLLD Riemann solver \citep{matsumoto2019}. A partial list of MHD codes built up with modern upwind schemes includes those by \citet{brio1988} using the Roe scheme, \citet{dai1994a,dai1994b} using the piecewise parabolic method (PPM), \citet{ryu1995a} and \citet{ryu1995b} using the total variation diminishing (TVD) scheme, and \citet{jiang1999} using the weighted essentially non-oscillatory (WENO) scheme. While these are designed to work on Eulerian grids, there are also codes that have been developed based on Lagrangian approaches, such as the smoothed particle hydrodynamics (SPH) MHD code \citep{price2012} and the moving-mesh MHD code \citep{mocz2016}.

We have developed a new code, HOW-MHD, High-Order WENO-based MHD, that solves ideal MHD equations with either adiabatic or isothermal EoS with high-order accuracy.\footnote{\textcolor{black}{The publicly accessible version of the HOW-MHD code can be found on \href {https://jeongbhin.github.io/} { https://jeongbhin.github.io/}.}} It includes the high-order finite-difference (FD) WENO scheme for the calculation of numerical fluxes and the high-order strong stability-preserving Runge–Kutta (SSPRK) method for time integration. Most of all, the code implements a new high-order version of the constrained transport (CT) scheme.

The WENO scheme is one of upwind schemes for solving hyperbolic conservation laws. It reconstructs fluxes with adaptive weights that depend on the smoothness indicators, and achieves a high-order accuracy in smooth flows and the non-oscillatory property near shocks and discontinuities \citep[see][for a review]{shu2009}. \citet{liu1994} first introduced weight functions for the third-order finite volume (FV) WENO scheme. And \citet{jiang1996} constructed third and fifth-order accurate weight functions for the FD WENO scheme, which has been widely used for studies of hydrodynamic processes. Since then, several WENO schemes with modified weight functions have been suggested, such as WENO-M \citep{henrick2005}, WENO-Z \citep{borges2008}, WENO-CU \citep{hu2010}, WENO-NS \citep{ha2013}, WENO-ZA \citep{liu2018}, and WENO-NIP \citep{li2022}. And \citet{jiang1999} built an  MHD code based on the WENO scheme by \citet{jiang1996}. We tested different WENO schemes, and found that \textcolor{black}{the fifth-order WENO version in \citet{jiang1996} and \citet{jiang1999}} works satisfactorily and is hence adopted for our MHD code.

Along with the WENO scheme, high-order Runge-Kutta (RK) methods have been commonly employed for time integration \citep[e.g.,][]{shu1988,shu1989,jiang1996}. While the combination of WENO and RK works well, some recent codes have been built with an improved SSPRK method \citep{spiteri2002,spiteri2003,gottlieb2005}. With SSPRK, spurious oscillations and smearing near discontinuous structures are reduced, and better results are obtained for complex flows \citep[e.g.,][]{christlieb2014}. In addition, SSPRK allows for a larger Courant-Friedrichs-Levy number, CFL $\geq1$, which leads to an improvement in computational efficiency. We adopt the fourth-order, five-stage version of SSPRK for our MHD code. CFL $\approx1.5$ is used as the default value.

In MHDs, the divergence-free constraint ($\mbox{\boldmath$\nabla$}\cdot\mbox{\boldmath$B$}=0$) needs to be maintained. While the MHD equations themselves formally comply with the constraint once it is initially satisfied, numerical errors arising from discretization and operator splitting can cause nonzero $\mbox{\boldmath$\nabla$}\cdot\mbox{\boldmath$B$}$ in multidimensional flows. Nonzero $\mbox{\boldmath$\nabla$}\cdot\mbox{\boldmath$B$}$ usually grows and eventually affects the correctness of flow dynamics \citep[see, e.g.,][]{brackbill1980}. Several methods have been proposed to enforce $\mbox{\boldmath$\nabla$}\cdot\mbox{\boldmath$B$}=0$, such as the vector potential approach \citep[e.g.,][]{clarke1986}, \textcolor{black}{the divergence cleaning method \citep[e.g.,][]{ryu1995b,dedner2002}}, the eight-wave method \citep{powell1999}, and the so-called CT scheme.

Of these methods, the CT scheme has become a popular approach. It was first introduced by \citet{evans1988}, and used in a number of upwind-based MHD codes \citep[e.g,][]{dai1998,ryu1998,balsara1999,toth2000,gardiner2008}. While the details of the algorithms in different codes somewhat vary \citep[see][for discussion on differences in different CT algorithms]{toth2000}, they all successfully keep $\mbox{\boldmath$\nabla$}\cdot\mbox{\boldmath$B$}=0$. The CT algorithms currently used in MHD codes typically employ second-order interpolation and FD. As a consequence, they can limit the overall accuracy, when they are used in codes based on high-order upwind schemes. For instance, \citet{donnert2019} showed that the MHD code based on the fifth-order WENO scheme produces second-order accuracy in convergence tests involving the damping of MHD waves in three-dimensional (3D) space, due to the second-order character of the CT part.

There have been efforts to preserve the high-order accuracy of MHD codes, by matching the order of the $\mbox{\boldmath$\nabla$}\cdot\mbox{\boldmath$B$}=0$ part. For instance, \citet{christlieb2014} used a high-order scheme to solve the vector potential, and \citet{minoshima2019} introduced a CT algorithm where a high-order FD is applied to compute the staggered magnetic field. For our MHD code, we have developed a new CT algorithm that generalizes the algorithm of \citet{ryu1998} to higher orders. The new algorithm, which updates the staggered magnetic field at grid cell interfaces using the advective fluxes, or effectively the electric field components, at grid cell edges, employs high-order interpolation and high-order FD while preserving $\mbox{\boldmath$\nabla$}\cdot\mbox{\boldmath$B$}=0$. We implement the new CT algorithm into our MHD code, completing its high-order competence.

In this paper, we present the Cartesian version of our MHD code, which includes the fifth-order WENO, the fourth-order, five-stage SSPRK, and the new high-order CT algorithm. Both the adiabatic and isothermal codes are described. We then present extensive tests to demonstrate the high accuracy and robustness of the code, including convergence tests with 3D MHD waves. We note that an early adiabatic version of the MHD code was described in \citet{donnert2019}, which includes the fifth-order WENO, the classical fourth-order RK4, and the second-order accurate CT algorithm of \citet{ryu1998}.

The paper is organized as follows. Section \ref{s2} includes the description of the code. Tests are presented in Section  \ref{s3}. A short summary follows in Section  \ref{s4}.

\section{Code Description}
\label{s2}

\subsection{Basic Equations}
\label{s2.1}

The equations for ideal adiabatic and isothermal MHDs can be found in the literature \citep[e.g.,][]{ryu1995a,kim1999}. We give them in the Appendix for completeness. Building upwind-scheme-based codes requires the eigenvalues and eigenvectors of characteristic modes. The eigenvalues are also listed in the Appendix, while we refer to the literature for the eigenvectors, whose expressions are rather long.  Below, $\rho$, \mbox{\boldmath$v$}, \mbox{\boldmath$B$}, $P$, and $\gamma$ are the fluid quantities, the density, velocity, magnetic field, pressure, and adiabatic index, respectively. And \mbox{\boldmath$q$} denotes the state vector, and \mbox{\boldmath$F$}, \mbox{\boldmath$G$}, and \mbox{\boldmath$H$} denote the flux vectors along the $x$-, $y$-, and $z$-directions, respectively, in the conservative form of MHD equations.

\subsection{FD WENO Scheme}
\label{s2.2}

The code is designed to update the state vector $\mbox{\boldmath$q$}_{i,j,k}$, defined at the centers of 3D Cartesian grid cells, with the dimension-by-dimension method as
\begin{equation}
\begin{aligned}
\mbox{\boldmath$q$}_{i,j,k}^{n+1} = \mbox{\boldmath$q$}_{i,j,k}^{n}
- \frac{\Delta t}{\Delta x}\left(\mbox{\boldmath$F$}^{n}_{i+\frac{1}{2},j,k}-\mbox{\boldmath$F$}^{n}_{i-\frac{1}{2},j,k}\right) ~~~~~~~~~~~\\
- \frac{\Delta t}{\Delta y}\left(\mbox{\boldmath$G$}^{n}_{i,j+\frac{1}{2},k}-\mbox{\boldmath$G$}^{n}_{i,j-\frac{1}{2},k}\right)
- \frac{\Delta t}{\Delta z}\left(\mbox{\boldmath$H$}^{n}_{i,j,k+\frac{1}{2}}-\mbox{\boldmath$H$}^{n}_{i,j,k-\frac{1}{2}}\right),\label{eq:qupdate}
\end{aligned}
\end{equation}
where the superscript $n$ indicates the time step, the subscripts $i$, $j$, and $k$ mark the spatial grid cells along the $x$-, $y$-, and $z$-directions, $\Delta x$, $\Delta y$, and $\Delta z$ are the cell sizes in the three directions, and $\Delta t$ is the duration between the $n$ and $n+1$ time steps. For estimating the numerical fluxes assigned at grid cell interfaces, $\mbox{\boldmath$F$}_{i\pm\frac{1}{2},j,k}$, $\mbox{\boldmath$G$}_{i,j\pm\frac{1}{2},k}$, and $\mbox{\boldmath$H$}_{i,j,k\pm\frac{1}{2}}$, \textcolor{black}{the fifth-order accurate FD WENO scheme is used \citep{jiang1996, jiang1999}.} \textcolor{black}{This WENO scheme is identical to that used in \citet{donnert2019}.}

We here brief the reconstruction of the $x$-flux $\mbox{\boldmath$F$}_{i+\frac{1}{2},j,k}$ with a stencil of the point-value flux $\{\mbox{\boldmath$F$}_i\}$ given at grid cell centers. The reconstruction of the $y$- and $z$-fluxes can be done by alternating the coordinates. In the rest of this subsection, we drop the subscripts $j$ and $k$ for simplicity. $\mbox{\boldmath$F$}_{i+\frac{1}{2}}$ is calculated as
\begin{equation}
\begin{aligned}
\mbox{\boldmath$F$}_{i+\frac{1}{2}}=\frac{1}{12}\left(-\mbox{\boldmath$F$}_{i-1}+7\mbox{\boldmath$F$}_{i}+7\mbox{\boldmath$F$}_{i+1}-\mbox{\boldmath$F$}_{i+2}\right) ~~ \\
+\sum_{s=1}^{7~{\rm or}~6}\Big[- \mathbf{\varphi}\left(\Delta\mbox{\boldmath$F$}^{s+}_{i-\frac{3}{2}},\Delta\mbox{\boldmath$F$}^{s+}_{i-\frac{1}{2}},\Delta\mbox{\boldmath$F$}^{s+}_{i+\frac{1}{2}},\Delta \mbox{\boldmath$F$}^{s+}_{i+\frac{3}{2}}\right) \\
+ \mathbf{\varphi}\left(\Delta\mbox{\boldmath$F$}^{s-}_{i+\frac{5}{2}},\Delta\mbox{\boldmath$F$}^{s-}_{i+\frac{3}{2}},\Delta\mbox{\boldmath$F$}^{s-}_{i+\frac{1}{2}},\Delta \mbox{\boldmath$F$}^{s-}_{i-\frac{1}{2}}\right)\Big]\mbox{\boldmath$R$}^{s}_{i+\frac{1}{2}},\label{5thweno}
\end{aligned}
\end{equation}
where $s$ denotes the seven (or six) characteristic modes of adiabatic (or isothermal) MHDs, and $\mbox{\boldmath$R$}^{s}_{i+\frac{1}{2}}$ is the right eigenvector. The characteristic modes are obtained using the left eigenvector $\mbox{\boldmath$L$}^{s}_{i+\frac{1}{2}}$ as
\begin{equation}
\mbox{\boldmath$F$}^{s}_{m}=\mbox{\boldmath$L$}^s_{i+\frac{1}{2}}\mbox{\boldmath$F$}_{m}, ~~~~~
\mbox{\boldmath$q$}^{s}_{m}=\mbox{\boldmath$L$}^s_{i+\frac{1}{2}}\mbox{\boldmath$q$}_{m},
\end{equation}
and their differences are
\begin{equation}
\Delta\mbox{\boldmath$F$}^{s}_{m+\frac{1}{2}}=\mbox{\boldmath$F$}^{s}_{m+1}-\mbox{\boldmath$F$}^{s}_{m}, ~~
\Delta\mbox{\boldmath$q$}^{s}_{m+\frac{1}{2}}=\mbox{\boldmath$q$}^{s}_{m+1}-\mbox{\boldmath$q$}^{s}_{m}.
\end{equation}
The local Lax-Friedrichs flux splitting is used to get $\Delta\mbox{\boldmath$F$}^{s\pm}_{m+\frac{1}{2}}$ for $m=i-2, \dots, i+2$:
\begin{equation}
\Delta\mbox{\boldmath$F$}^{s\pm}_{m+\frac{1}{2}} = \frac{1}{2}\left(\Delta\mbox{\boldmath$F$}^{s}_{m+\frac{1}{2}}\pm\mathbf{\delta}_{i+\frac{1}{2}}^{s}\Delta\mbox{\boldmath$q$}^{s}_{m+\frac{1}{2}}\right),
\end{equation}
where $\mathbf{\delta}_{i+\frac{1}{2}}^{s}=\max(|\lambda^s_m|)$ is the maximum of the $s$th eigenvalues within $i-2\leq m\leq i+3$. The eigenvalues are given in the Appendix. For the left and right eigenvectors, we use those  given in \citet{ryu1995a} and \citet{kim1999}; for the calculation of the eigenvectors at grid cell interfaces, we use the arithmetic averaging of fluid quantities at the centers of two neighboring grid cells.

The interpolant function $\mathbf{\varphi}$ is defined as
\begin{equation}
\begin{aligned}
\mathbf{\varphi}(a_1,a_2,a_3,a_4) = \frac{1}{3}\omega_{0}(a_1-2a_2+a_3)\\
+\frac{1}{6}\left(\omega_{2}-\frac{1}{2}\right)(a_2-2a_3+a_4).~~~~~
\end{aligned}
\end{equation}
Here, $\omega_0$ and $\omega_2$ are the weight functions, which are given as
\begin{equation}
\omega_r = \frac{\alpha_{r}}{\sum_{r'=0}^2\alpha_{r'}},~~\alpha_{r} = \left(\frac{\mathcal{C}_{r}}{\epsilon+IS_{r}}\right)^2, ~~ r=0,1,2,
\end{equation}
where $\mathcal{C}_{0}=0.1$, $\mathcal{C}_{1}=0.6$, and $\mathcal{C}_{2}=0.3$, respectively. The local smoothness indicators $IS_{r}$ are given as
\begin{equation}
\begin{aligned}
IS_{0} = 13(a_1-a_2)^{2}+3(a_1-3a_2)^{2},\\
IS_{1} = 13(a_2-a_3)^{2}+3(a_2+a_3)^{2},~\\
IS_{2} = 13(a_3-a_4)^{2}+3(3a_3-a_4)^{2}.
\end{aligned}
\end{equation}
The parameter $\epsilon$ is included to avoid the zero denominator, and $\epsilon=10^{-8}$ is used.

\subsection{SSPRK Time Integration}
\label{s2.3}

For the time advance of the state vector, the fourth-order, five-stage SSPRK method is adopted \citep{spiteri2002,spiteri2003,gottlieb2005}. The state vector at the time step $n$ is updated to $n+1$ as follows:
\begin{equation}
\begin{aligned}
\mbox{\boldmath$q$}_{i,j,k}^{(0)}=\mbox{\boldmath$q$}_{i,j,k}^{n}, ~~~~~~~~~~~~~~~~~~~~~~~~~~~~~~~~~~~~~~~~~\\
\mbox{\boldmath$q$}_{i,j,k}^{(l)}=\sum_{m=0}^{l-1}(\chi_{lm} \mbox{\boldmath$q$}_{i,j,k}^{(m)}+\Delta t\beta_{lm} \mbox{\boldmath$\cal L$}_{i,j,k}^{(m)}),~~~~~~~~~~\\
l = 1,2,\cdots,5,~~~~~~~~~~~~~~~~~~~~~~~~~~~~~~~~\\
\mbox{\boldmath$q$}_{i,j,k}^{n+1}=\mbox{\boldmath$q$}_{i,j,k}^{(5)}. ~~~~~~~~~~~~~~~~~~~~~~~~~~~~~~~~~~~~~~~~~\label{eq:SSPRK}
\end{aligned}
\end{equation}
Here, $\mbox{\boldmath$\cal L$}^{(l)}_{i,j,k}$ is given as
\begin{equation}
\begin{aligned}
\mbox{\boldmath$\cal L$}^{(l)}_{i,j,k} = -\frac{\mbox{\boldmath$F$}^{(l)}_{i+\frac{1}{2},j,k}-\mbox{\boldmath$F$}^{(l)}_{i-\frac{1}{2},j,k}}{\Delta x} ~~~~~~~~~~~~\\
-\frac{\mbox{\boldmath$G$}^{(l)}_{i,j+\frac{1}{2},k}-\mbox{\boldmath$G$}^{(l)}_{i,j-\frac{1}{2},k}}{\Delta y}
-\frac{\mbox{\boldmath$H$}^{(l)}_{i,j,k+\frac{1}{2}}-\mbox{\boldmath$H$}^{(l)}_{i,j,k-\frac{1}{2}}}{\Delta z},
\label{eq:SSPRK2}
\end{aligned}
\end{equation}
where $\mbox{\boldmath$\cal L$}^{(l)}_{i,j,k}$ is calculated with $\mbox{\boldmath$q$}_{i,j,k}^{(l)}$. The coefficients $\chi_{lm}$ and $\beta_{lm}$ are given in \citet{spiteri2002}.

For the numerical stability, the time step, $\Delta t$, is restricted by the CFL condition:
\begin{equation}
\Delta t = {\rm CFL}/\left[\frac{\lambda^{\rm max}_{x}}{\Delta x}+\frac{\lambda^{\rm max}_{y}}{\Delta y}+\frac{\lambda^{\rm max}_{z}}{\Delta z}\right],
\end{equation}
where $\lambda^{\rm max}$'s are the maxima of the eigenvalues at grid cell centers, $\lambda_{i,j,k}^s$, in the entire computational domain, along the $x$-, $y$-, and $z$-directions, respectively. Normally, CFL $<1$ is required. However, SSPRK allows CFL $>1$; according to \citet{spiteri2002}, the optimal value of CFL is given as $\min (\chi_{lm}/\beta_{lm})$, which is 1.50818004975927 for the fourth-order, five-stage SSPRK. We use it as the default value in our code. We point out that the fourth-order, five-stage SSPRK with CFL $\approx1.5$ results in a $\sim50\%$ increase in computational efficiency, compared to the classical fourth-order, four-stage RK4 typically with CFL=0.8.

\begin{figure*}
\vskip 0.1 cm
\hskip 0.7 cm
\includegraphics[width=0.9\linewidth]{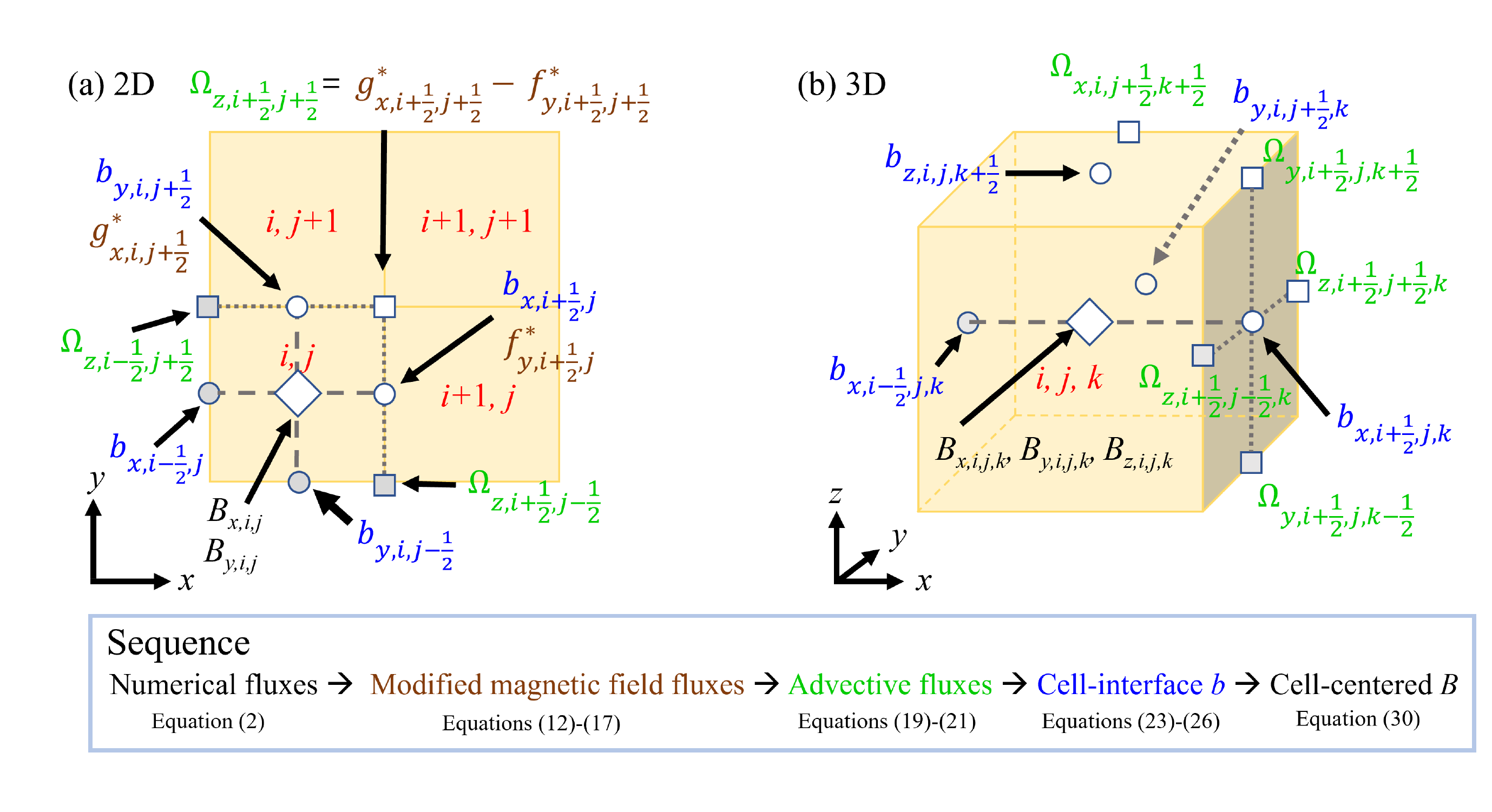}
\vskip -0.5 cm
\caption{Schematic pictures displaying the quantities involved in our CT algorithm in (a) 2D and (b) 3D geometries. While the fluid variables in the state vector, including the magnetic field $\mbox{\boldmath$B$}$, are given at grid cell centers, the divergence-free magnetic field, $\mbox{\boldmath$b$}$, is defined at grid cell interfaces, and the advective flux, $\mbox{\boldmath$\Omega$}$, is defined at grid cell edges. The sequence of the CT algorithm is summarized in the bottom box.}
\label{f1}
\end{figure*}

\subsection{High-order CT}
\label{s2.4}

To match the spatial accuracy of the fifth-order WENO scheme, our code incorporates a newly developed high-order CT algorithm that complies with the divergence-free constraint. The new CT algorithm builds upon the second-order accurate CT algorithm of \citet{ryu1998}; it is based on the staggered approach, where the components of the divergence-free magnetic field, $\mbox{\boldmath$b$}$, are defined at grid cell interfaces, and updated using the advective fluxes evaluated at grid cell edges, $\mbox{\boldmath$\Omega$}$ (see Figure \ref{f1}), via the induction equation for magnetic field evolution. To achieve high-order accuracy, the scheme employs high-order interpolation and high-order FD.

For the new CT algorithm, the ``modified magnetic field fluxes'' is introduced: 
\textcolor{black}{\begin{eqnarray}
f^*_{y,i+\frac{1}{2},j,k} = f_{y,i+\frac{1}{2},j,k} + B_{x,i+\frac{1}{2},j,k}v_{y,i+\frac{1}{2},j,k}\label{mmff1}\\
f^*_{z,i+\frac{1}{2},j,k} = f_{z,i+\frac{1}{2},j,k} + B_{x,i+\frac{1}{2},j,k}v_{z,i+\frac{1}{2},j,k}\label{mmff2}\\
g^*_{z,i,j+\frac{1}{2},k} = g_{z,i,j+\frac{1}{2},k} + B_{y,i,j+\frac{1}{2},k}v_{z,i,j+\frac{1}{2},k}\\
g^*_{x,i,j+\frac{1}{2},k} = g_{x,i,j+\frac{1}{2},k} + B_{y,i,j+\frac{1}{2},k}v_{x,i,j+\frac{1}{2},k}\\
h^*_{x,i,j,k+\frac{1}{2}} = h_{x,i,j,k+\frac{1}{2}} + B_{z,i,j,k+\frac{1}{2}}v_{x,i,j,k+\frac{1}{2}}\\
h^*_{y,i,j,k+\frac{1}{2}} = h_{y,i,j,k+\frac{1}{2}} + B_{z,i,j,k+\frac{1}{2}}v_{y,i,j,k+\frac{1}{2}} 
\end{eqnarray}}
where $f$, $g$, and $h$ are the WENO-reconstructed magnetic field fluxes along the $x$-, $y$-, and $z$-directions at grid cell interfaces, respectively, and $B_{x,i+\frac{1}{2},j,k}v_{y,i+\frac{1}{2},j,k}$ and others are the quantities at grid cell interfaces, interpolated with $B_{x,i,j,k}v_{y,i,j,k}$ at grid cell centers. As the fifth-order WENO reconstruction is adopted, the optimal interpolation is fourth-order accurate: for an arbitrary function, $\mathcal{A}_{i,j,k}$, the fourth-order interpolation along the $x$-direction is given as
\begin{equation}
\mathcal{A}_{i+\frac{1}{2},j,k} =
\frac{1}{16}(-\mathcal{A}_{i-1,j,k}+9\mathcal{A}_{i,j,k} +9\mathcal{A}_{i+1,j,k}-\mathcal{A}_{i+2,j,k}), \label{4int}
\end{equation}
and the interpolations along the $y$- and $z$-directions are given similarly. The modified magnetic field fluxes intend to include the contributions of the upwinding terms only; for instance, those in Equations (\ref{mmff1}) and (\ref{mmff2}) basically contain the fluxes due to the $B_y v_x$ and $B_z v_x$ terms in the sixth and seventh columns of Equations (\ref{MHDflux1}) and (\ref{MHDflux2}). The contributions of the other $B_x v_y$ and $B_x v_z$ terms are counted in the calculation of the advective fluxes below.

Then, the advective fluxes, or effectively the electric field components, at grid cell edges are given as
\begin{eqnarray}
\Omega_{z,i+\frac{1}{2},j+\frac{1}{2},k} = g^*_{x,i+\frac{1}{2},j+\frac{1}{2},k}-f^*_{y,i+\frac{1}{2},j+\frac{1}{2},k},\label{omegaz}\\
\Omega_{x,i,j+\frac{1}{2},k+\frac{1}{2}} = h^*_{y,i,j+\frac{1}{2},k+\frac{1}{2}}-g^*_{z,i,j+\frac{1}{2},k+\frac{1}{2}},\label{omegax}\\
\Omega_{y,i+\frac{1}{2},j,k+\frac{1}{2}} = f^*_{z,i+\frac{1}{2},j,k+\frac{1}{2}}-h^*_{x,i+\frac{1}{2},j,k+\frac{1}{2}}\label{omegay}.
\end{eqnarray}
Here, the modified magnetic field fluxes at grid cell edges are calculated with those at grid cell interfaces, again using the fourth-order interpolation; for instance, $g^*_{x,i+\frac{1}{2},j+\frac{1}{2},k}$ is calculated with $g^*_{x,i,j+\frac{1}{2},k}$ using the same formula as Equation (\ref{4int}).

\textcolor{black}{The update of the magnetic field components at grid cell interfaces ($b_{x,i+\frac{1}{2},j,k}$, $b_{y,i,j+\frac{1}{2},k}$, and $b_{z,i,j,k+\frac{1}{2}}$ in Figure \ref{f1}) is done through the derivatives of the advective fluxes. In the WENO scheme, for the derivatives of fluxes along a specific direction, high order is achieved with two-point FD using high-order reconstructed fluxes along the direction of the derivative. For example, to calculate the first term on the right-hand side of Equation (\ref{eq:qupdate}), the flux reconstructed to high order along the $x$-direction, $\mbox{\boldmath$F$}_{i+\frac{1}{2},j,k}$, is used. On the other hand, each of the above advective fluxes includes two fluxes that are reconstructed along two directions, such as $g^*_{x,i+\frac{1}{2},j+\frac{1}{2},k}$and $f^*_{y,i+\frac{1}{2},j+\frac{1}{2},k}$ in Equation (\ref{omegaz}). Hence, the straightforward application of two-point FD, or even multi-point high-order FD (e.g., in Equation (\ref{FD}), below), does not necessarily result in high-order accuracy in CT.}

\textcolor{black}{To achieve high-order accuracy in our CT algorithm, we implement the following two steps for the update of the magnetic field components. First, the advective fluxes are modified to approximate ``point values'' at grid cell edges:} for the advective flux $\Omega_{z,i+\frac{1}{2},j+\frac{1}{2},k}$ at the $x-y$ edges of grid cells,
\begin{equation}
\bar{\Omega}_{z,i+\frac{1}{2},j+\frac{1}{2},k} = \frac{1}{\Delta x\Delta y}\int_{\sigma_{xy}}\Omega_{z,i+\frac{1}{2},j+\frac{1}{2},k}dxdy - {\cal O}(\Delta^{\mu}).
\end{equation}
\textcolor{black}{Here, $\sigma_{xy}=[x_{i},x_{i+1}]\times[y_{i},y_{i+1}]$ is in the plane defined with the reconstruction directions of involved fluxes, $g^*_{x,i+\frac{1}{2},j+\frac{1}{2},k}$ and $f^*_{y,i+\frac{1}{2},j+\frac{1}{2},k}$,} and ${\cal O}(\Delta^\mu)$ denotes the error of the order of accuracy. Again, the fourth-order accuracy would be optimal, \textcolor{black}{which is expressed as \citep{buchmuller2014}}
\begin{equation}
\begin{aligned}
\bar{\Omega}_{z,i+\frac{1}{2},j+\frac{1}{2},k} = {\Omega}_{z,i+\frac{1}{2},j+\frac{1}{2},k}~~~~~~~~~~~~~~\\
+\frac{1}{24}\left({\Omega}_{z,i-\frac{1}{2},j+\frac{1}{2},k}-2{\Omega}_{z,i+\frac{1}{2},j+\frac{1}{2},k}+{\Omega}_{z,i+\frac{3}{2},j+\frac{1}{2},k}\right)~\\
+\frac{1}{24}\left({\Omega}_{z,i+\frac{1}{2},j-\frac{1}{2},k}-2{\Omega}_{z,i+\frac{1}{2},j+\frac{1}{2},k}+{\Omega}_{z,i+\frac{1}{2},j+\frac{3}{2},k}\right).\label{baromega}
\end{aligned}
\end{equation}
For the other advective flux components, $\bar{\Omega}_{x,i,j+\frac{1}{2},k+\frac{1}{2}}$ and $\bar{\Omega}_{y,i+\frac{1}{2},j,k+\frac{1}{2}}$ are calculated in the similar way.

\textcolor{black}{Second}, a high-order FD is applied to the derivatives of the above \textcolor{black}{point-value fluxes \citep[e.g.,][]{delzanna2007}} as
\begin{eqnarray}
b_{x,i+\frac{1}{2},j,k}^{n+1}=b_{x,i+\frac{1}{2},j,k}^n - \frac{\Delta t}{\Delta y}\mathcal{D}_{\nu,y}(\bar{\Omega}_z) + \frac{\Delta t}{\Delta z}\mathcal{D}_{\nu,z}(\bar{\Omega}_y),~~~~\\
b_{y,i,j+\frac{1}{2},k}^{n+1}=b_{y,i,j+\frac{1}{2},k}^n - \frac{\Delta t}{\Delta z}\mathcal{D}_{\nu,z}(\bar{\Omega}_x) + \frac{\Delta t}{\Delta x}\mathcal{D}_{\nu,x}(\bar{\Omega}_z),~~~~\\
b_{z,i,j,k+\frac{1}{2}}^{n+1}=b_{x,i,j,k+\frac{1}{2}}^n - \frac{\Delta t}{\Delta x}\mathcal{D}_{\nu,x}(\bar{\Omega}_y) + \frac{\Delta t}{\Delta y}\mathcal{D}_{\nu,y}(\bar{\Omega}_x),~~~~
\end{eqnarray}
where $\mathcal{D}_\nu$ denotes the FD operator of the order of accuracy $\nu$. To maintain the overall spatial accuracy of the code at fifth order, the FD of fifth or higher order would be necessary. We adopt sixth-order accurate FD as the default: for an arbitrary function, $\mathcal{A}_{i,j,k}$, the sixth-order FD along the $x$-direction is given as
\begin{equation}
\begin{aligned}
\mathcal{D}_{\nu,x}(\mathcal{A}_{i\pm})_i=c_1(\mathcal{A}_{i+\frac{1}{2}}-\mathcal{A}_{i-\frac{1}{2}})~~~~~~~\\
+~c_2(\mathcal{A}_{i+\frac{3}{2}}-\mathcal{A}_{i-\frac{3}{2}})+c_3(\mathcal{A}_{i+\frac{5}{2}}-\mathcal{A}_{i-\frac{5}{2}}),
\end{aligned} \label{FD}
\end{equation}
where $c_1=75/64$, $c_2=-25/384$, and $c_3=3/640$. Here, the subscripts $j$ and $k$ are dropped for simplicity. The FDs along the $y$- and $z$-directions are given similarly. To be comprehensive, fourth-order accurate FD is also considered, for which $c_1=9/8$, $c_2=-1/24$, and $c_3=0$ in the above equation. \textcolor{black}{Below, the CT algorithms using the above sixth- and fourth-order FDs are referred to as CT6 and CT4, respectively.} Interestingly, we have found that with both CT6 and CT4, the code achieves overall fifth-order spatial accuracy in convergence tests with 3D MHD waves and produces comparable results for complex flows, as shown in the next section.

The magnetic field components at grid cell interfaces are updated with the SSPRK steps: for $b_{x,i+\frac{1}{2},j,k}$
\begin{equation}
\begin{aligned}
b_{x,i+\frac{1}{2},j,k}^{(0)}=b_{x,i+\frac{1}{2},j,k}^{n},~~~~~~~~~~~~~~~~~~~~~~~~~~~~~~~~~~~~~~~~~~~~~~\\
b_{x,i+\frac{1}{2},j,k}^{(l)}=\sum_{m=0}^{l-1}\chi_{lm} b_{x,i+\frac{1}{2},j,k}^{(m)}~~~~~~~~~~~~~~~~~~~~~~~~~~~~~~~~~~~~\\
- \frac{\Delta t}{\Delta y}\beta_{lm} \mathcal{D}_{\nu,y}(\bar{\Omega}_{z,i+\frac{1}{2},j\pm,k}^{(m)})_j
+ \frac{\Delta t}{\Delta z}\beta_{lm} \mathcal{D}_{\nu,z}(\bar{\Omega}_{y,i+\frac{1}{2},j,k\pm}^{(m)})_k,\\
l = 1,2,\cdots,5,~~~~~~~~~~~~~~~~~~~~~~~~~~~~~~~~~~~~~~~~\\
b_{x,i+\frac{1}{2},j,k}^{n+1}=b_{x,i+\frac{1}{2},j,k}^{(5)},~~~~~~~~~~~~~~~~~~~~~~~~~~~~~~~~~~~~~~~~~~~~~~
\end{aligned}
\end{equation}
and similarly for $b_{y,i,j+\frac{1}{2},k}$ and $b_{z,i,j,k+\frac{1}{2}}$.

With the FD of the order of accuracy $\nu$, the divergence of $\mbox{\boldmath$b$}$ can be calculated as
\begin{equation}
\begin{aligned}
(\mbox{\boldmath$\nabla$}\cdot\mbox{\boldmath$b$})_{i,j,k} = \frac{1}{\Delta x}\mathcal{D}_{\nu,x}(b_{x,i\pm,j,k})_{i}~~~~~~\\
+ \frac{1}{\Delta y}\mathcal{D}_{\nu,y}(b_{y,i,j\pm,k})_{j} + \frac{1}{\Delta z}\mathcal{D}_{\nu,z}(b_{z,i,j,k\pm})_{k}.
\end{aligned} \label{divbFD}
\end{equation}
Then, the divergence-free constraint, $\mbox{\boldmath$\nabla$}\cdot\mbox{\boldmath$b$}=0$, is exactly satisfied, up to the numerical truncation error (see Figure \ref{f10} below).

Finally, the magnetic field in the state vector at grid cell centers, $\mbox{\boldmath$B$}_{i,j,k}$, is calculated by interpolating the magnetic field at grid cell interfaces. Again, to maintain overall spatial accuracy at fifth order, the interpolation needs to be fifth- or higher-order accurate. We adopt sixth-order interpolation: for $B_{x,i,j,k}$, the interpolation along the $x$-direction is given as
\begin{equation}
\begin{aligned}
B_{x,i,j,k} = ~~~~~~~~~~~~~~~~~~~~~~~~~~~~~~~~~~~~~~~~~~~~~~\\
\frac{1}{256}(3b_{x,i-\frac{5}{2},j,k}-25b_{x,i-\frac{3}{2},j,k}+150b_{x,i-\frac{1}{2},j,k}\\
+150b_{x,i+\frac{1}{2},j,k}-25b_{x,i+\frac{3}{2},j,k}+3b_{x,i+\frac{5}{2},j,k}), \label{stateB}
\end{aligned}
\end{equation}
and for $B_{y,i,j,k}$ and $B_{z,i,j,k}$, the interpolations are given similarly. We have found that lower-order interpolations result in lower-order convergence in tests with 3D MHD waves; for instance, with fourth-order interpolation, only fourth-order accuracy is obtained.

We note that high-order FD was previously applied to the discretization of the induction equation, for instance, in the MHD code presented in \citet{minoshima2019}. However, in different codes, the advective fluxes, or effectively the electric field components, are estimated differently. In addition, the interpolations involved are done differently. In our CT algorithm, the estimation of the advective fluxes in Equation (\ref{baromega}), the FD in Equation (\ref{FD}), and the calculation of $\mbox{\boldmath$B$}_{i,j,k}$ in Equation (\ref{stateB}) are optimized in such a way as to preserve the spatial accuracy of the WENO scheme in multi-dimensional problems.

\begin{figure}
\vskip 0.1 cm
\hskip -0.1 cm
\includegraphics[width=1\linewidth]{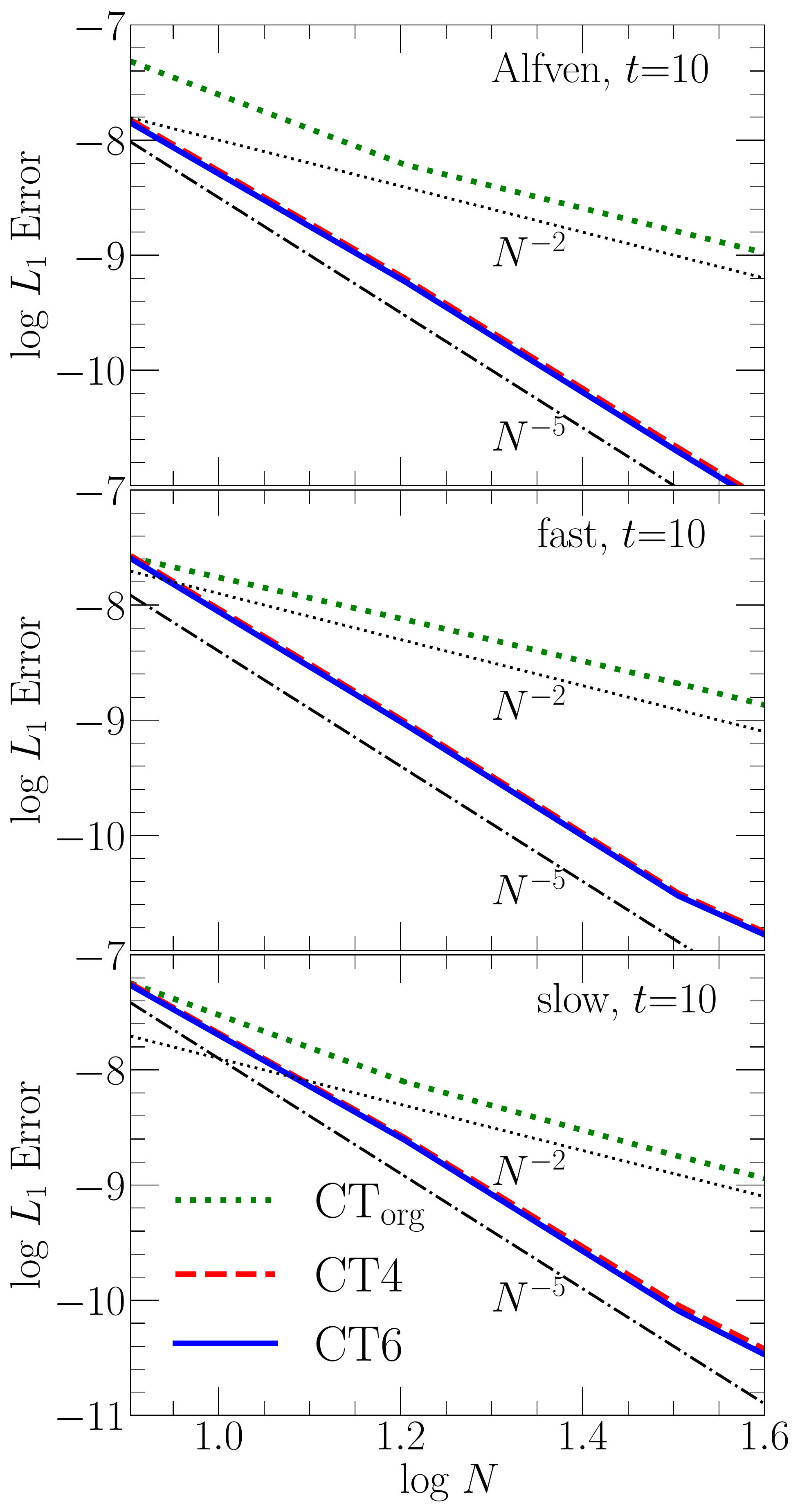}
\vskip -0.1 cm
\caption{$L_1$ errors due to the damping of three linear MHD waves - Alfv\'en, fast, and slow - at different resolutions. The waves propagate in 3D space at an oblique angle to the coordinate axes. Here and in the following figures, the results with the CT algorithm employing sixth-order FD ($\nu=6$ in Equation (\ref{FD})) and fourth-order FD ($\nu=4$ in Equation (\ref{FD})) are labeled as CT6 and CT4, respectively, and the results with the CT algorithm of \citet{ryu1998} are labeled as CT$_{\rm org}$. The lines showing the fifth- and second-order convergences are drawn for comparison.}
\label{f2}
\end{figure}

\begin{figure}
\vskip 0.1 cm
\hskip -0.1 cm
\includegraphics[width=1\linewidth]{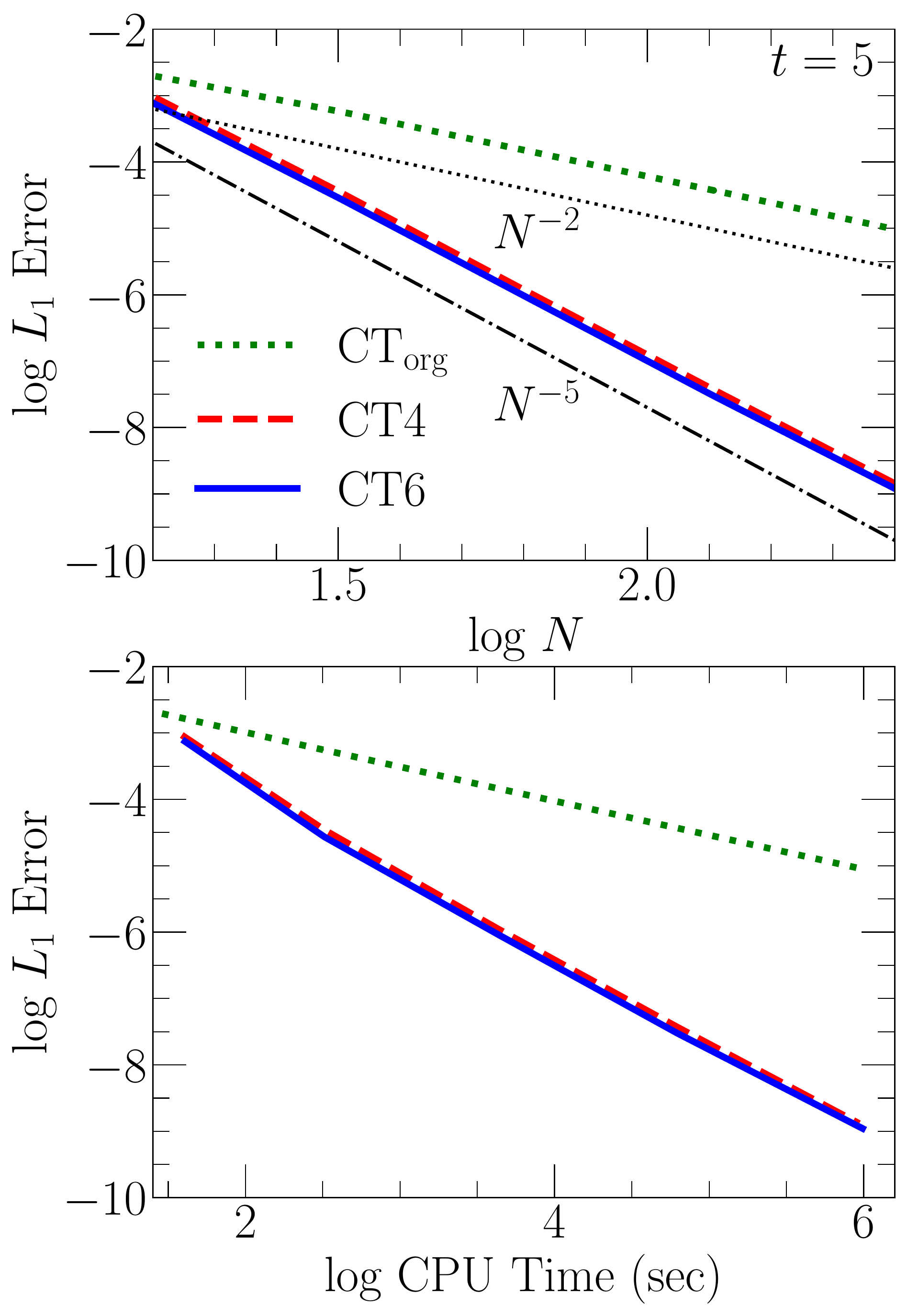}
\vskip -0.1 cm
\caption{Top panel: $L_1$ errors at different resolutions due to the damping of a circularly polarized Alfv\'en wave propagating in 3D space at an oblique angle to the coordinate axes. The lines showing the fifth- and second-order convergences are drawn for comparison. \textcolor{black}{Bottom panel: $L_1$ errors vs CPU time on a computer with twenty cores.}}
\label{f3}
\end{figure}

\section{Code Verification Tests}
\label{s3}

A series of tests have been conducted to establish the high accuracy and robustness of our MHD code. We present some of them in three categories. Tests in the first category (in the first four subsections) intend to verify the order of accuracy of the code and also highlight the improvements with the new high-order CT algorithm. Tests in the second category (in the next four subsections) demonstrate the reliability and robustness of the code along with its high accuracy, for problems involving shocks and complex flows. The tests for these two categories have been done using the adiabatic MHD code. Tests in the third category (in the final two subsections) include simulations of turbulent flows with both the isothermal and adiabatic MHD codes to verify the performance of the isothermal code and also to compare the results of the two codes.

CFL $\approx1.5$ is used in all the tests presented; $\gamma=5/3$ is used in the tests of adiabatic flows, except in the MHD rotor test, where $\gamma=1.4$ is used following previous works (see Section \ref{s3.6}). The primitive variables of the initial MHD state are denoted as $\mbox{\boldmath$u$} = (\rho, v_{x},v_{y},v_{z},B_{x},B_{y},B_{z}, p)^T$.

\begin{figure*}
\centering
\vskip 0.1 cm
\includegraphics[width=0.85\linewidth]{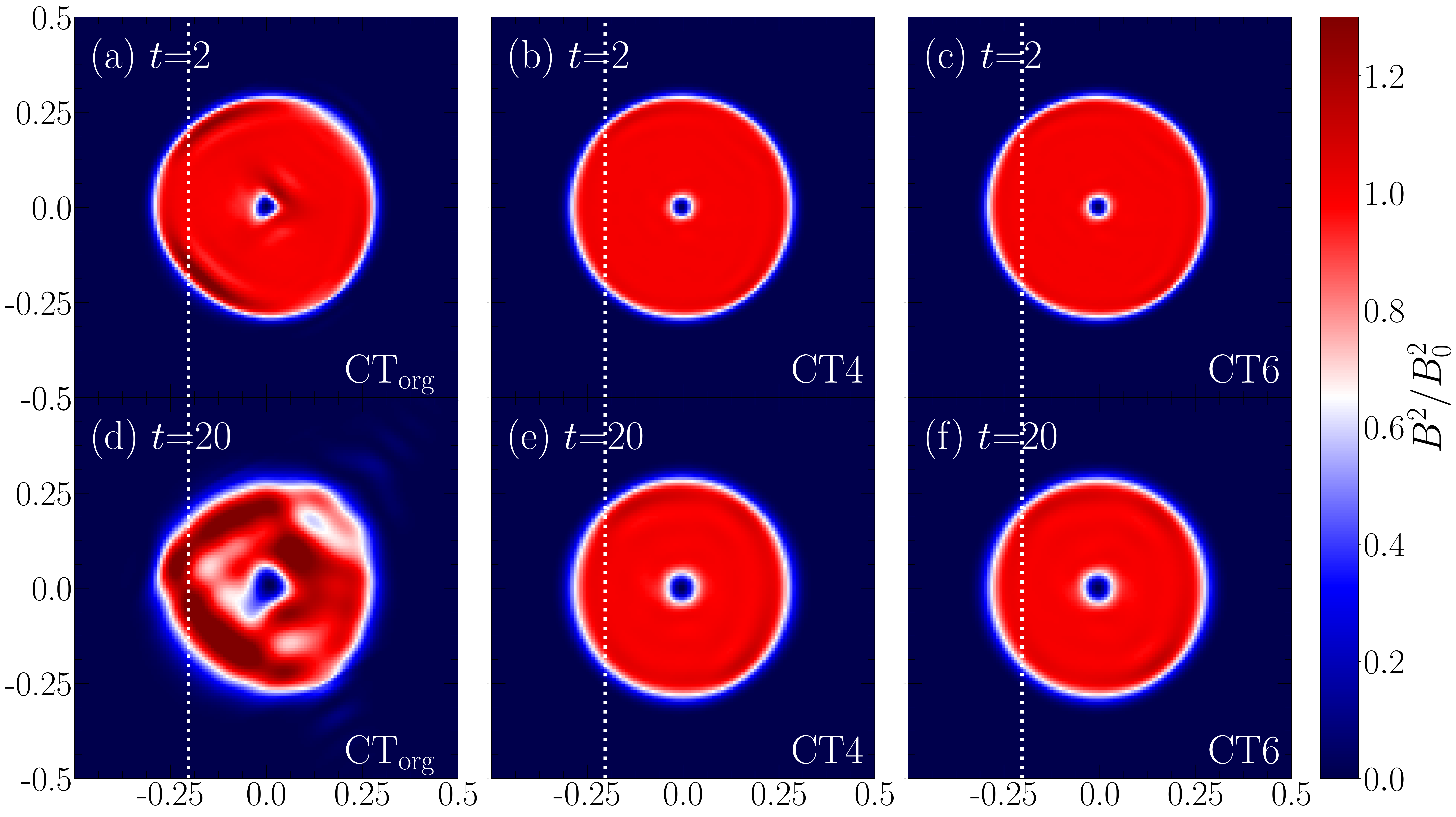}
\vskip 0 cm
\caption{Normalized magnetic energy, $B^2/B_0^2$, of the circular magnetic loop advecting across a diagonal direction in the elongated 2D domain of $[-1,1]\times[-1/2,1/2]$ with $256\times128$ grid cells. The test results using the codes with CT6, CT4, and CT$_{\rm org}$ are shown after advecting the computational domain twice ($t=2$, top panels) and twenty times  ($t=20$, bottom panels). The profiles of $B^2/B_0^2$ along the white dotted lines are shown in the top panel of Figure \ref{f5}.}
\label{f4}
\end{figure*}

\begin{figure}
\vskip 0.1 cm
\hskip -0.1 cm
\includegraphics[width=1\linewidth]{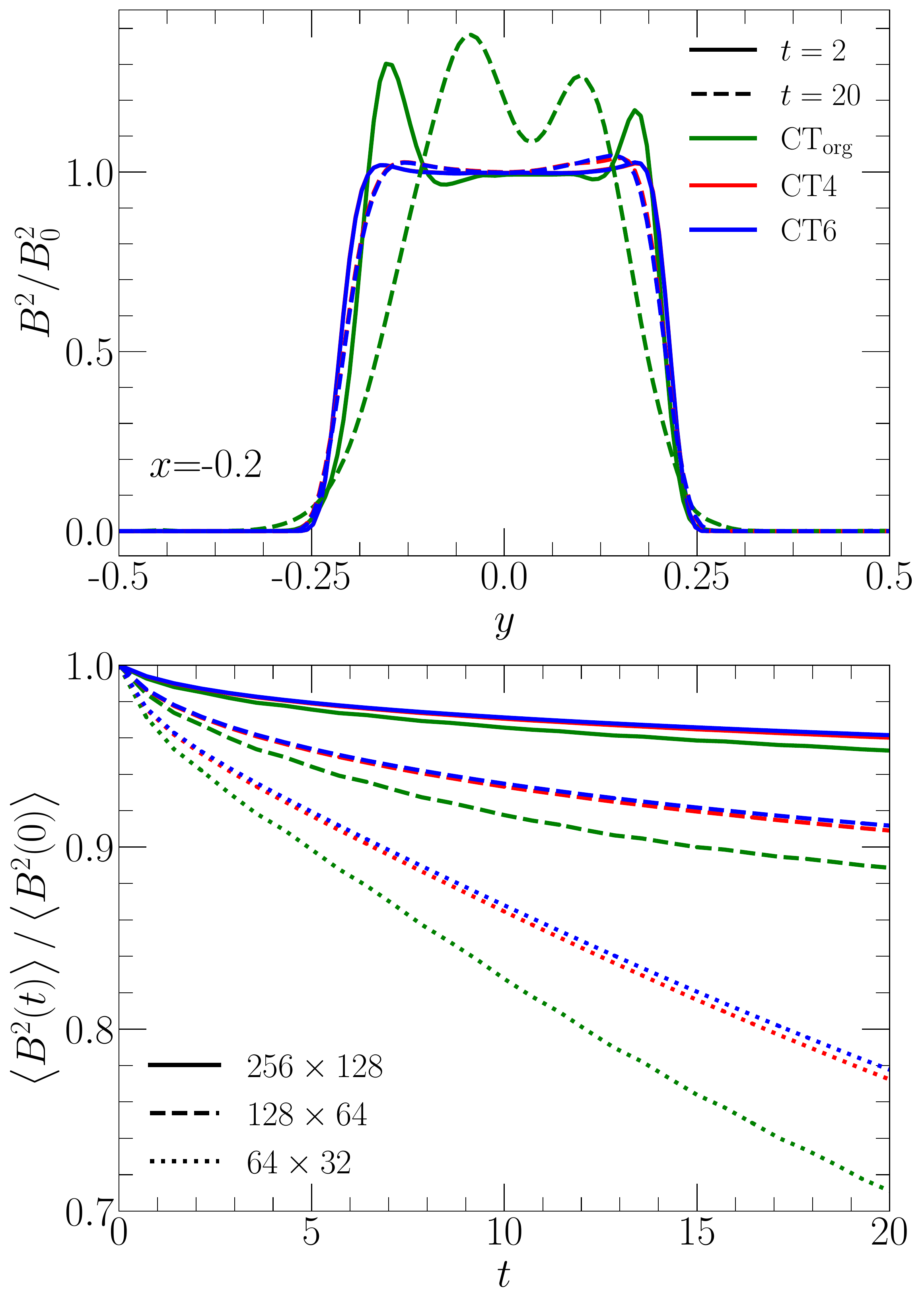}
\vskip -0.1 cm
\caption{Top panel: Normalized magnetic energy, $B^2/B_0^2$, of the circular magnetic loops advecting across the 2D  computational domain, along the white dotted lines shown in Figure \ref{f4}. Bottom panel: Time evolution of the average of the normalized magnetic energy of the 2D circular magnetic loops in test simulations with $256\times128$, $128\times64$, and $64\times32$ grid cells.}
\label{f5}
\end{figure}

\subsection{Convergence Test Using Linear MHD Waves Propagating in 3D}
\label{s3.1}

We evaluate the convergence order of the code using the damping of linear MHD waves at different resolutions. The damping should be the consequence of numerical diffusivity. Following previous works \citep{gardiner2008,stone2008,donnert2019}, we consider linear MHD waves propagating in 3D space at an oblique angle to the coordinate axes.

Linear MHD waves along the $x$-direction are given as
\begin{equation}
\mbox{\boldmath$u$}=\bar{\mbox{\boldmath$u$}}+A_0\delta\mbox{\boldmath$u$}\sin(2\pi x),
\end{equation}
where the unperturbed state is
\begin{equation}
\bar{\mbox{\boldmath$u$}}=(1,0,0,0,1,\sqrt{2},1/2,1/\gamma)^T, \\
\end{equation}
and the perturbed part is the right-hand eigenvectors of the modes of three MHD waves - Alfv\'en, fast, and slow,
\begin{eqnarray}
\delta\mbox{\boldmath$u$}_{\rm{Alfven}}=\frac{1}{6\sqrt{5}}(0,0,1,-2\sqrt{2},0,-1,2\sqrt{2},0)^T, \\
\delta\mbox{\boldmath$u$}_{\rm{fast}}=\frac{1}{6\sqrt{5}}(6,12,-4\sqrt{2},-2,0,8\sqrt{2},4,27)^T, \\
\delta\mbox{\boldmath$u$}_{\rm{slow}}=\frac{1}{6\sqrt{5}}(12,6,8\sqrt{2},4,0,-4\sqrt{2},-2,9)^T.
\end{eqnarray}
For the amplitude of perturbation, $A_0=10^{-6}$ is used. The MHD waves tested are set up by rotating the above waves with the Euler angles of $-\arctan(2/\sqrt{5})$ and $\arctan(2)$ about the $y$- and $z$-axes \citep[see][for details]{gardiner2008}.

Test simulations are run in a 3D periodic box of $3\times3/2\times3/2$ volume with $2N\times N\times N$ grid cells. Then, the $L_1$ error due to the damping of the waves is estimated with the conserved variables of the state vector $\mbox{\boldmath$q$}$ as
\textcolor{black}{\begin{equation}
\sqrt{\sum_{p=1}^{8}\left( \sum_{i,j,k} \frac{|\mbox{\boldmath$q$}^p_{i,j,k}(t)-\mbox{\boldmath$q$}^p_{i,j,k}(0)|}{2N^3}\right)^2},
\label{l1error}
\end{equation}}
where the inner summation covers the entire computational domain and \textcolor{black}{the outer summation with $p$ is over the eight conserved variables}.

Figure \ref{f2} plots the $L_1$ errors for the three linear MHD waves in simulations with resolutions of $N=8,~16,~32$, and 64 at $t=10$; the results obtained with the new CT algorithm are labeled as CT6 (sixth-order FD in Equation (\ref{FD})) and CT4 (fourth-order FD in Equation (\ref{FD})), while those with the CT algorithm of \citet{ryu1998} are labeled as CT$_{\rm org}$. As noted in the introduction, in \citet{donnert2019}, the code with CT$_{\rm org}$ achieved second-order convergence due to the second nature of CT$_{\rm org}$, even though the code was built with the WENO scheme of fifth-order accuracy. We also observe that our code produces second-order convergence with CT$_{\rm org}$. On the other hand, with the new CT algorithms, both CT6 and CT4, the code achieves fifth-order convergence, successfully preserving the order of accuracy of WENO. Furthermore, while the convergence order is the same, CT6 yields slightly smaller $L_1$ errors compared to CT4. Considering that the additional cost of CT6 is negligible, we choose CT6 as the fault scheme, as stated in Section \ref{s2.4}.

\subsection{Convergence Test Using a Circularly Polarized Alfv\'en Wave Propagating in 3D}
\label{s3.2}

\begin{figure}
\centering
\vskip 0.1 cm
\hskip -0.6 cm
\includegraphics[width=1.07\linewidth]{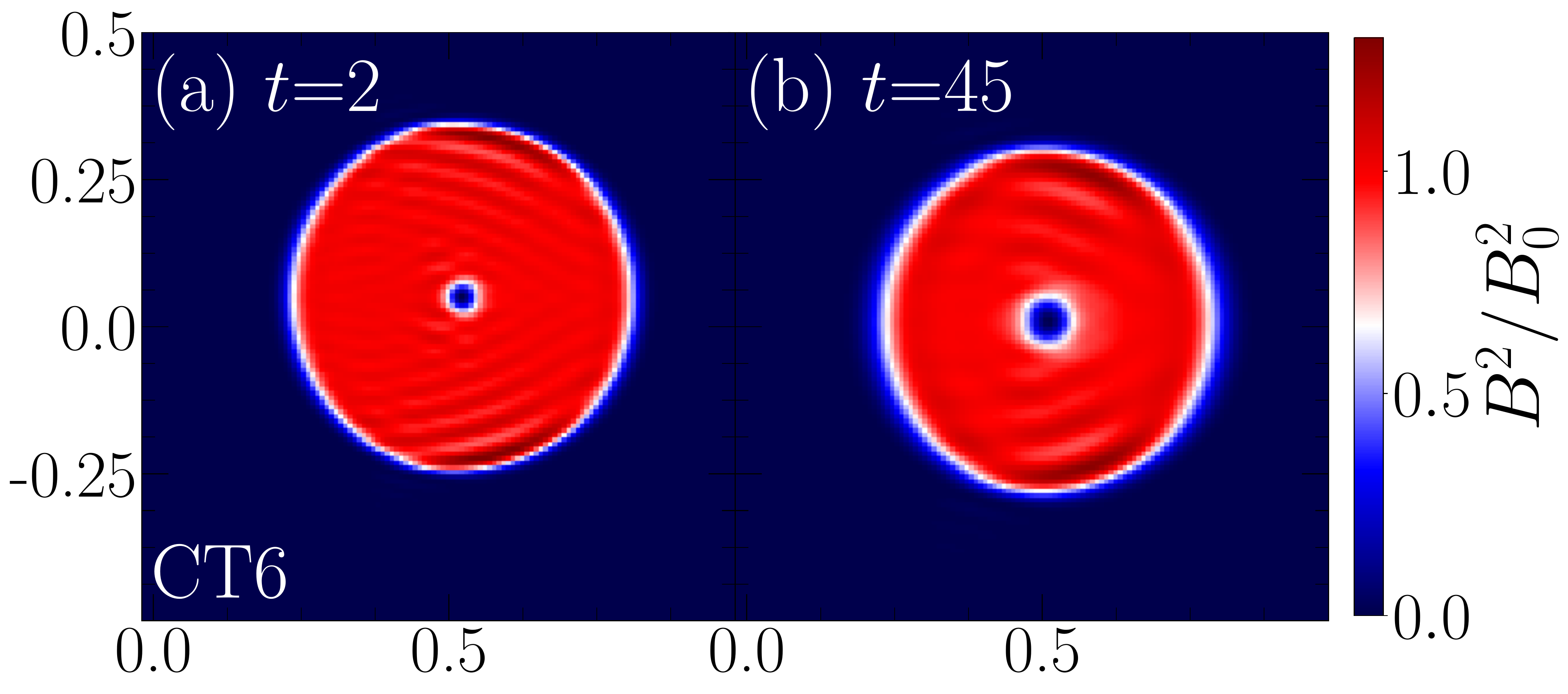}
\vskip -0.1 cm
\caption{\textcolor{black}{Normalized magnetic energy, $B^2/B_0^2$, of the circular magnetic loop advecting with a small angle of $\tan\theta=1/100$ in the 2D domain of $[-1,1]\times[-1/2,1/2]$ with $256\times128$ grid cells. Here, $\theta$ is the advection angle with respect to the $x$-axis. The test results using the codes with CT6 are shown at $t=2$ (a) and $t=45$ (b).}}
\label{f6}
\end{figure}

\begin{figure*}
\centering
\vskip 0.1 cm
\includegraphics[width=0.9\linewidth]{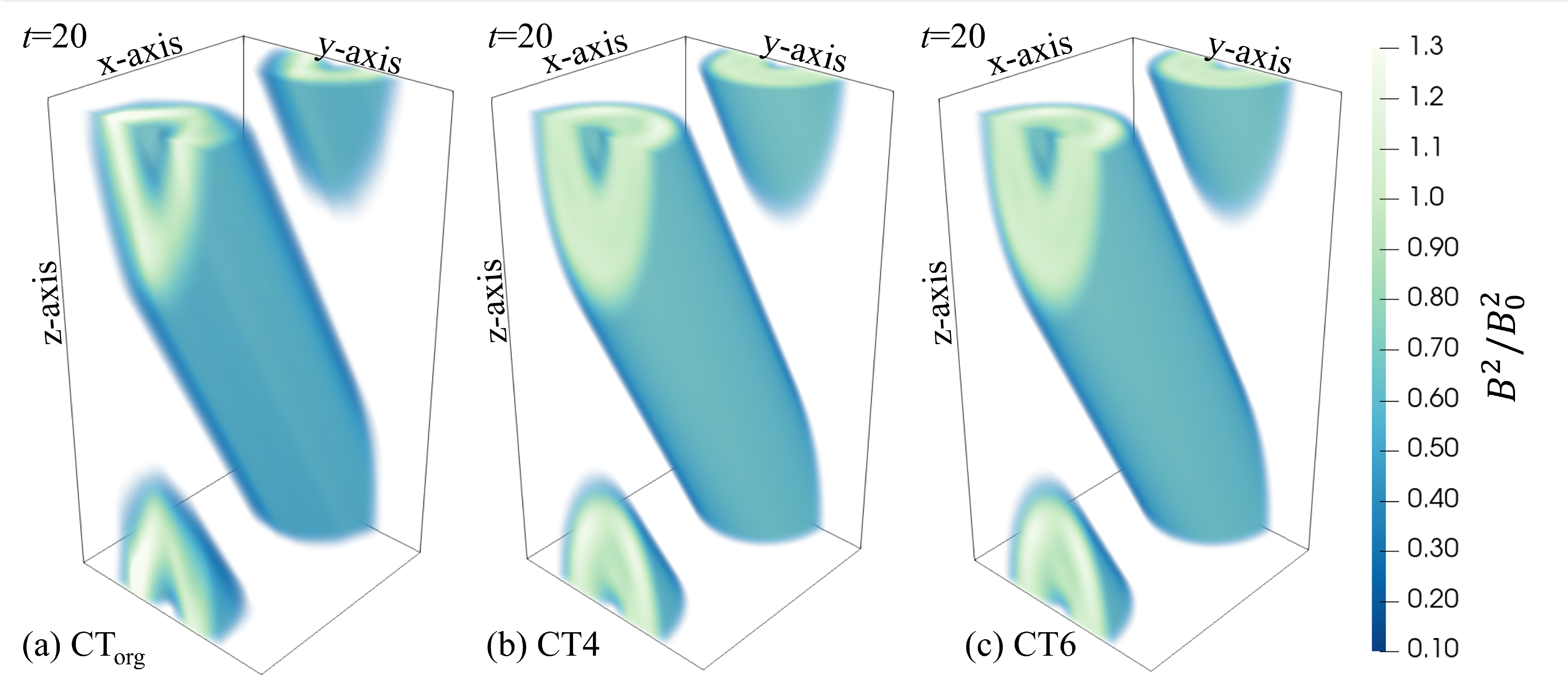}
\vskip -0.1 cm
\caption{Volume rendering images of $B^2/B_0^2$, the normalized magnetic energy of the inclined, cylindrical magnetic loop advecting across a diagonal direction in the 3D domain of $1\times1\times2$ volume with $64\times64\times128$ grids. The test results using the codes with CT6, CT4, and CT$_{\rm org}$ at $t=20$ are displayed.}
\label{f7}
\end{figure*}

Polarized Alfv\'en waves are commonly found in astrophysical environments such as the solar corona \citep[e.g.,][]{goldstein1978}. Hence, we verify the convergence order of the code again with a circularly polarized Alfv\'en waves. Following previous works \citep{gardiner2008,stone2008,donnert2019}, we consider a wave propagating in 3D space at an oblique angle: the wave along the $x$-direction is given as
\begin{equation}
\mbox{\boldmath$u$}=(1,0,v_y,v_z,1,B_y,B_z,0.1)^T,\\
\end{equation}
with $v_y=B_y=0.1\sin(2\pi x)$, $v_z=B_z=0.1\cos(2\pi x)$, and the tested wave is set up by rotating it with the Euler angles of $-\arctan(2/\sqrt{5})$ and $\arctan(2)$ about the $y$- and $z$-axes.

Simulations are run in a 3D periodic box of $3\times3/2\times3/2$ volume with $2N\times N\times N$ grid cells.\textcolor{black}{The top panel of} Figure \ref{f3} plots the $L_1$ errors (Equation (\ref{l1error})) in simulations with resolutions of $N=16,~32,~64,~128$, and 256 at $t=5$. Again, with CT$_{\rm org}$, second-order convergence is obtained, as in \citet{donnert2019}. In contrast, with CT6 and CT4, the code achieves fifth-order convergence, and with CT6, the $L_1$ error is slightly smaller. \textcolor{black}{In the bottom panel of Figure \ref{f3}, the CPU efficiency in the test is shown. It demonstrates that the computational times are not much different among the CT schemes, and hence CT6 and CT4 yield higher accuracy per time.}

\subsection{Advection of a Magnetic Field Loop in 2D and 3D}
\label{s3.3}

The advection of a magnetic field loop was introduced as a test problem for MHD codes in \citet{gardiner2005} and subsequently presented in a number of papers \citep[e.g.,][]{gardiner2008,stone2008,donnert2019,minoshima2019,mignone2021}. It has turned out to be a simple yet not trivial test; the loop is distorted, and the distortion is sensitive to the accuracy and numerical diffusivity of the code. We here present the 2D and 3D versions, where a circular (2D) or cylindrical (3D) magnetic field loop moves across a diagonal direction.

In the 2D test, the initial state is given as
\textcolor{black}{\begin{equation}
\mbox{\boldmath$u$}=(1,u_0\cos\theta,u_0\sin\theta,v_z,B_x, B_y, 0, 1)^T,
\end{equation}
where $\theta$ is the advection angle with respect to the $x$-axis.} The magnetic field at grid cell interfaces is set up as $\mbox{\boldmath$b$}=\mbox{\boldmath$\nabla$}\times\mbox{\boldmath$A$}$ with the vector potential assigned at grid cell edges,
\begin{equation}
A_z=
\begin{cases}
A_0(r_c-r)&{\rm for~}r\leq r_c \\
0~~~~~~~~~~~ & {\rm for~}r>r_c
\end{cases}, \label{advloop}
\end{equation}
where $r=\sqrt{x^2+y^2}$, $r_c = 0.3$, and $A_0=10^{-3}$ \citep{gardiner2005}. The magnetic field at grid cell centers, $\mbox{\boldmath$B$}$, are calculated using Equation (\ref{stateB}). The computational domain consists of $[-1,1]\times[-1/2,1/2]$, covered with $2N\times N$ grid cells, in the $x-y$ plane, and the boundaries are periodic. Simulations are run using the codes with CT6, CT4, and CT$_{\rm org}$.

\textcolor{black}{Figure \ref{f4} displays the results of simulations with $u_0=\sqrt{5}$, $\tan\theta=1/2$, and $v_z=0$ using $256\times128$ grid cells;} the distributions of the normalized magnetic energy of the loop, $B^2/B_0^2$, after advecting across the computational domain twice ($t=2$, top panels) and twenty times ($t=20$, bottom panels) are shown. With CT$_{\rm org}$, the loop becomes clearly distorted over time. On the contrary, with the new CT algorithm, both CT6 and CT4, the circular shape of the loop remains well-maintained. The top panel of Figure \ref{f5} draws the normalized magnetic energy in the loops along the $x=-0.2$ line shown in Figure \ref{f4}. Again, the improvement with the new CT algorithm is evident; the erosion of the loop reduces substantially, and the oscillations caused by the discontinuity at the loop perimeter disappear almost completely.

\begin{figure*}
\vskip 0.1 cm
\hskip 1.0 cm
\includegraphics[width=0.85\linewidth]{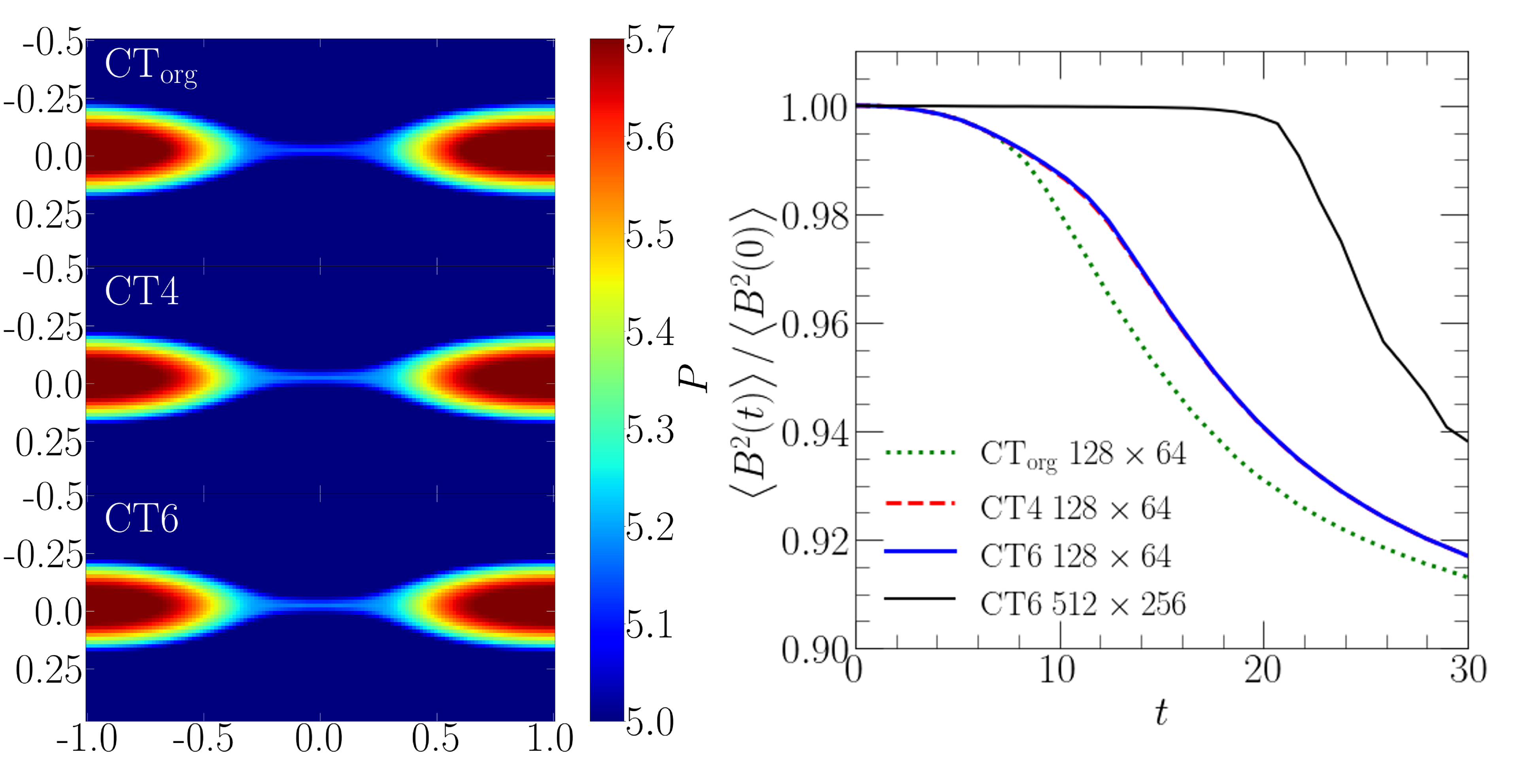}
\vskip -0.2 cm
\caption{Results of the 2D magnetic reconnection test in the $x-y$ domain of $[-1,1]\times[-0.5,0.5]$. See the main text for the setup of the current sheet. Left panel: The 2D distributions of the gas pressure, $P$, in simulations with $128\times64$ grid cells, using the codes with CT6, CT4, and CT$_{\rm org}$, at $t=30$. Right panel: The time evolution of the magnetic energy in the entire computational domain, normalized to the initial magnetic energy, in the same simulations shown in the left panel. The result of a higher resolution simulation with $512\times256$ grid cells is also shown for comparison.}
\label{f8}
\end{figure*}

The bottom panel of Figure \ref{f5} plots the time evolution of the average of the normalized magnetic energy in simulations with $N=32$, 64, and 128. The magnetic energy gradually decreases over time due to the numerical diffusivity of the code. The amount of the magnetic energy being retained is larger with CT6 and CT4 than with CT$_{\rm org}$; at $t=20$, $\sim96\%$ of the magnetic energy is retained with the new algorithm, whereas it is $\sim95\%$ with CT$_{\rm org}$, in simulations with $N=128$, and the difference is larger at lower resolutions. Moreover, the results with CT6 are better with slightly more retained energy than those with CT4.

\textcolor{black}{While the above is the test performed in most of the literature, it can be repeated with different advection angles to examine how numerical diffusivity operates in different directions. We here present the test with $u_0=\sqrt{5}$, $\tan\theta=1/100$ (hence, $v_x \gg v_y$), and $v_z=1$, which was presented in \citet{lee2013}. Figure \ref{f6} shows the distributions of $B^2/B_0^2$ at $t=2$ and $t=45$ in simulation using the code with CT6. With the advection mostly along the $x$-direction, the amount of numerical diffusivity in different directions should be different. As a consequence, spurious oscillations of amplitude up to a few percent appear along the $y$-direction. However, the oscillations do not grow over time. In addition, despite the oscillations, the circular shape of the loop is well maintained.}
 
For the 3D test, the magnetic loop used for the test shown in Figure \ref{f4} is stretched along the $z$-direction and tilted at an angle of $\arctan(1/2)=22.5$ degrees about the $y$-axis. In practice, the tilted cylindrical magnetic loop in 3D is set up using the 3D vector potential, $\mbox{\boldmath$A$}$, obtained by properly rotating the one in Equation (\ref{advloop}) \citep[see][for details]{donnert2019}; $\mbox{\boldmath$b$}$ is calculated with $\mbox{\boldmath$A$}$ at grid cell interfaces, and then $\mbox{\boldmath$B$}$ is calculated at grid cell centers. Simulations are run using the codes with CT6, CT4, and CT$_{\rm org}$, in a 3D periodic box of $1\times1\times2$ volume with $N\times N\times 2N$ grid cells.

Figure \ref{f7} displays the 3D distributions of the normalized magnetic energy of the loop in simulations with $N=64$ at $t=20$. Again, with the new algorithm, both CT6 and CT4, the cylindrical shape of the loop is well preserved. In contrast, with CT$_{\rm org}$, the erosion is apparent in the perimeter and core of the loop, and oscillatory features are observed inside the loop.

\begin{figure*}
\centering
\vskip 0.1 cm
\includegraphics[width=0.9\linewidth]{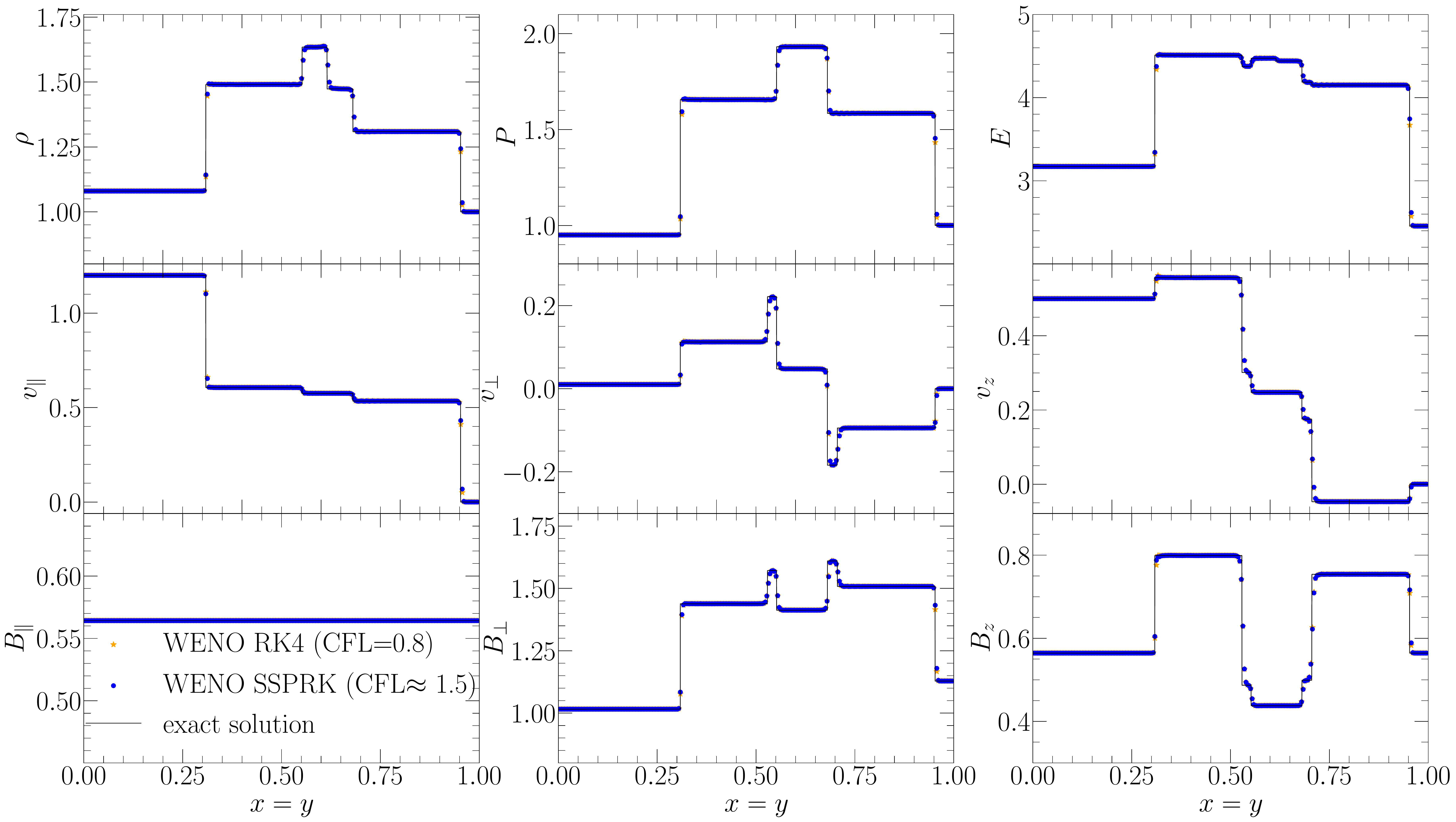}
\vskip -0.1 cm
\caption{2D oblique MHD shock-tube test, the same test shown in Figure 2 of \citet{ryu1995b}. See the main text for the setup. The computational domain covers the $x-y$ plane of $[0,1]\times[0,1]$ with $256\times256$ grid cells. The results of a simulation at $t=0.2\times\sqrt{2}$ using the code with SSPRK and CT6 are shown with blue dots. For comparison, the results using the code with RK4 and CT6 are also shown with orange stars. The numerical solutions are along the diagonal line of $x=y$, and compared to the exact solution of \citet{ryu1995a} drawn with solid lines.}
\label{f9}
\end{figure*}

\subsection{2D Magnetic Reconnection}
\label{s3.4}

We present a test involving the reconnection of magnetic fields, particularly the 2D reconnection test performed by \citet{mignone2021}. The initial state is given as
\begin{equation}
\mbox{\boldmath$u$}=(1,0,0,0,B_x+\delta B_x,\delta B_y,0,P)^T, 
\end{equation}
in the computational domain of $[-1,1]\times[-1/2,1/2]$, covered with $2N\times N$ grid cells. For the mean magnetic field, a Harris current sheet is adopted:
\begin{equation}
B_x(y) = B_0 \tanh(y/\delta_L)
\end{equation}
with $B_0=1$ and $\delta_L=0.04$. The gas pressure profile is adjusted to achieve an equilibrium configuration,
\begin{equation}
P(y) = \frac{B_0^2}{2}(\beta_p+1)-\frac{B_x(y)^2}{2}
\end{equation}
with $\beta_p=10$ which is the plasma beta at $y\rightarrow\infty$. To initiate the reconnection, small perturbations of the magnetic field are added: with the vector potential assigned at grid cell edges,
\begin{equation}
\delta A_z = 10^{-3} B_0\cos(2\pi x/L_x)\cos(\pi y/L_y),
\end{equation}
$\delta\mbox{\boldmath$b$}$ is calculated at grid cell interfaces, and then $\delta\mbox{\boldmath$B$}$ is calculated at grid cell centers. Here, $L_x=2$ and $L_y=1$ are the $x$- and $y$-sizes of the computational domain, respectively. Simulations are run using the codes with CT6, CT4, and CT$_{\rm org}$. Boundaries are periodic at the left and right sides and reflecting at top and bottom.

The left panel of Figure \ref{f8} depicts the distributions of the gas pressure, $P$, in simulations with $N=64$ at $t=30$. The distributions with CT6, CT4, and CT$_{\rm org}$ are visually alike, all displaying the typical shape of the current sheet region. The right panel of Figure \ref{f8} shows the time evolution of the normalized magnetic energy for three simulations with $N=64,$ and also that from a higher resolution simulation with CT6. The plot indicates that the reconnection proceeds more quickly with CT$_{\rm org}$ than with the new CT algorithm; the results with CT6 and CT4 are comparable and closer to the high-resolution results. In this test where simulations do not include physical resistivity, the reconnection is induced by numerical diffusivity. Hence, the test results tell us that the numerical diffusivity of the code effectively decreases with the new CT algorithm, compared to CT$_{\rm org}$.

\subsection{2D Oblique MHD Shock Tube}
\label{s3.5}

The ability of the code to capture shocks and discontinuities can be examined in shock-tube tests. We present a 2D oblique shock-tube test in the $x-y$ plane of $[0,1]\times[0,1]$, specifically the test shown in Figure 2 of \citet{ryu1995b} where two fast shocks, two rotational discontinuities, two slow shocks, and a contact discontinuity form and propagate to either the left or right direction along the diagonal line of $x=y$. The initial state is given in the left and right regions, separated  along $x+y=1$, with the left and right states,
\begin{equation}
\begin{aligned}
\mbox{\boldmath$u$}_{L}=\Big(1.08,\frac{v_{\parallel,L}-v_{\perp,L}}{\sqrt{2}},\frac{v_{\parallel,L}+v_{\perp,L}}{\sqrt{2}},0.5,~~~~~~~~~~~~\\
\frac{B_{\parallel}-B_{\perp,L}}{\sqrt{2}},\frac{B_{\parallel}+B_{\perp,L}}{\sqrt{2}},\frac{2}{\sqrt{4\pi}},0.95\Big)^T, \\
\mbox{\boldmath$u$}_{R} =\Big(1,0,0,0,\frac{B_{\parallel}-B_{\perp,R}}{\sqrt{2}},\frac{B_{\parallel}+B_{\perp,R}}{\sqrt{2}},\frac{2}{\sqrt{4\pi}},1\Big)^T,
\label{RJ2a}
\end{aligned}
\end{equation}
where $v_{\parallel,L}=1.2$, $v_{\perp,L}=0.01$, $B_{\parallel}=2/\sqrt{4\pi}$, $B_{\perp,L}=3.6/\sqrt{4\pi}$, , $B_{\perp,R}=4/\sqrt{4\pi}$. This is the case that the initial state with $v_x=v_{\parallel}$, $v_y=v_{\perp}$, $B_x=B_{\parallel}$, and $B_y=B_{\perp}$ is rotated at an angle of 45 degrees about the $z$-axis. In practice, we set up the initial magnetic field using the vector potential given as
\begin{equation}
A_z=
\begin{cases}
-(B_{\parallel}+B_{\perp,L})x/\sqrt{2}+(B_{\parallel}-B_{\perp,L})y/\sqrt{2} \\
~~~~~~~~~~~~~~~~~~~~~~~~~~~~~~~~~~~~~~~~~{\rm for~} x+y\leq1 \\
-(B_{\parallel}+B_{\perp,R})x/\sqrt{2}+(B_{\parallel}-B_{\perp,R})y/\sqrt{2} \\ 
~~~~~~~~~~~~~~~~~~~~~~~~~~~~~~~~~~~~~~~~~{\rm for~} x+y>1
\end{cases},
\end{equation}
at grid cell edges; $\mbox{\boldmath$b$}$ is calculated with $A_z$ at grid cell interfaces, and then $\mbox{\boldmath$B$}$ is calculated at grid cell centers.

\begin{figure*}
\centering
\vskip 0.1 cm
\includegraphics[width=0.9\linewidth]{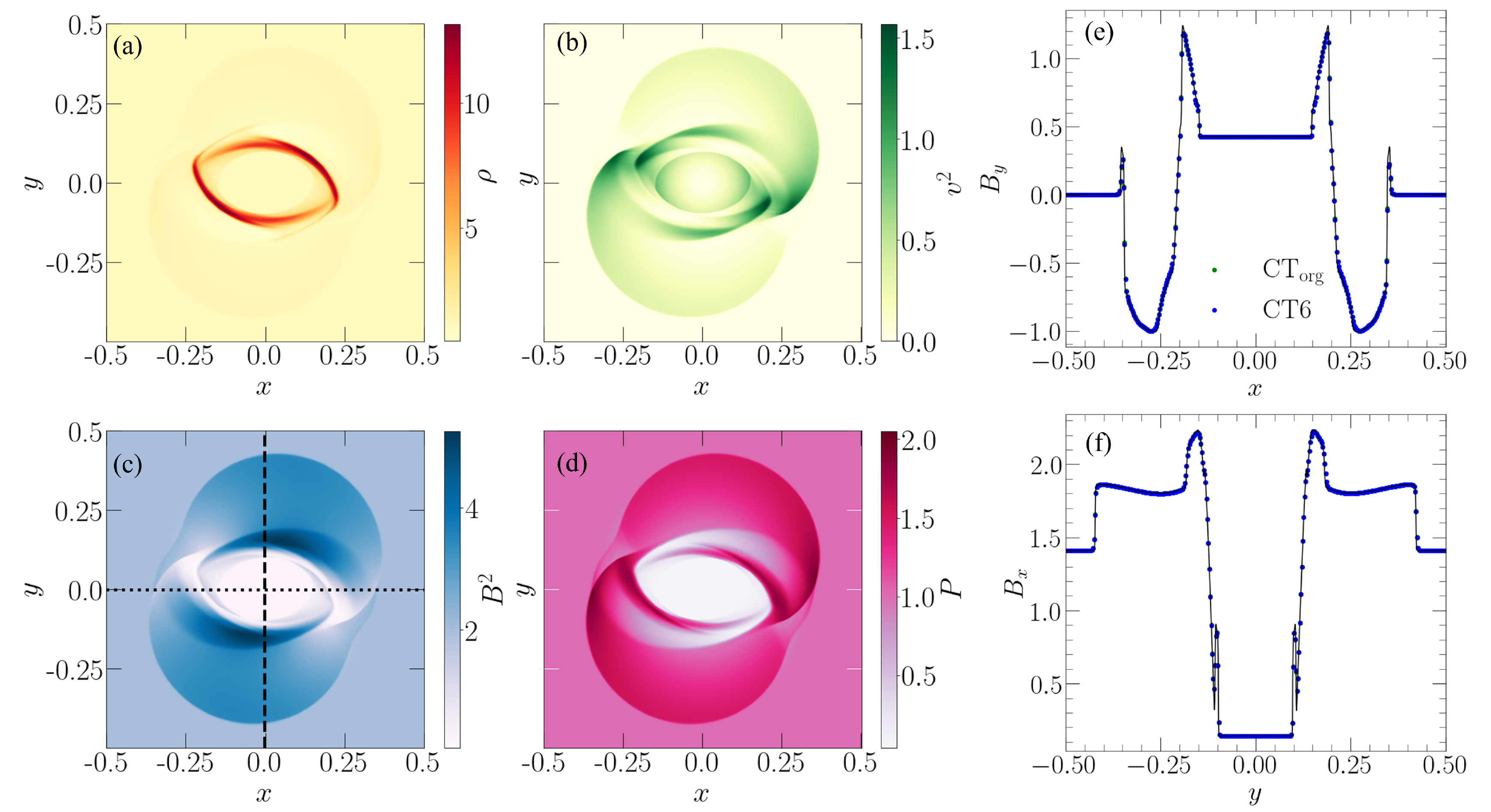}
\vskip -0.1 cm
\caption{Results of the 2D MHD rotor test. See the main text for the setup. The computational domain consists of $[-0.5,0.5]\times[-0.5,0.5]$ with $400\times400$ grid cells in the $x-y$ plane. Panels (a) to (d): The distributions of $\rho$, $v^2$, $B^2$, and $P$. The results of a simulation at $t=0.125$ using the code with CT6 are shown. Panels (e) and (f): The line profiles of $B_y$ along the $x$-axis ($y=0$, dotted line in panel (c)) and $B_x$ along the $y$-axis ($x=0$, dashed line in panel (c)). The results of simulations using the codes with CT6 and CT$_{\rm org}$ are shown. The solid lines draw the results of a higher resolution simulation with $1024\times1024$ grid cells using the code with CT6 for comparison.}
\label{f10}
\end{figure*}

\begin{figure}
\vskip 0.1 cm
\hskip -0.1cm
\includegraphics[width=1.\linewidth]{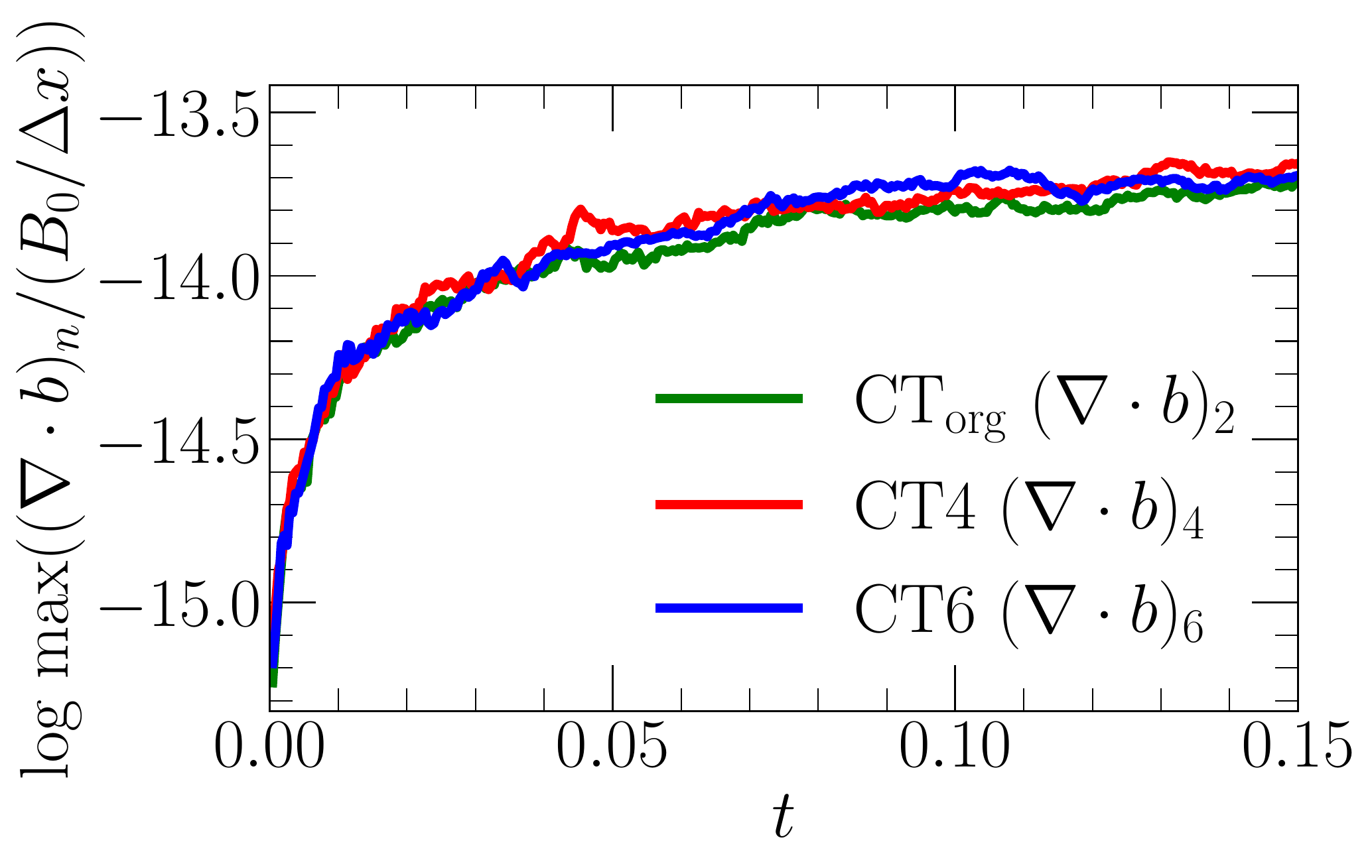}
\vskip -0.1 cm
\caption{Time evolution of the maximum value of $\mbox{\boldmath$\nabla$}\cdot\mbox{\boldmath$b$}$, normalized with the initial magnetic field and the cell size, $B_0/\Delta x$, in the 2D MHD rotor test with $400\times400$ grid cells using the codes with CT6, CT4, and CT$_{\rm org}$. The values of $\mbox{\boldmath$\nabla$}\cdot\mbox{\boldmath$b$}$ are calculated with Equation (\ref{divbFD}) using Equation (\ref{FD}) for the cases of CT6 ($\nu=6$) and CT4 ($\nu=4$) and using a second-order FD for CT$_{\rm org}$, matched with the order of the CT algorithms.}
\label{f11}
\end{figure}

Figure \ref{f9} shows the results of a simulation with $256\times256$ grid cells using the code with CT6 (blue dots), along with the exact analytic solution of \citet{ryu1995a} for comparison, at $t=0.2\sqrt{2}$. All the structures are correctly reproduced; shocks and discontinuities spread typically over two to three grid cells. Despite the fact that high-order interpolation and high-order FD are employed in our new CT algorithm, we do not observe any noticeable degradation in capturing shocks and discontinuities. The results with CT4 and CT$_{\rm org}$, although not shown here, are basically identical to those with CT6, indicating that the ability to capture shocks and discontinuities is not sensitive to the CT algorithm once the divergence-free condition is satisfied. On the contrary, our code built with the high-order WENO scheme does a somewhat better job than, for instance, the second-order TVD code, particularly in capturing rotational and contact discontinuities, as can be seen in this figure and Figure 2b of \citet{ryu1998}.

In Figure \ref{f9}, the results using the version of the code with RK4, instead of SSPRK, for time integration are also shown (orange stars); CFL $=0.8$ is used, and otherwise, the code is the same, including CT6. With most of the blue dots and the orange stars overlapping, the results with SSPRK and RK4 are basically identical. Both versions of the code successfully capture all the structures. On the other hand, with CFL $\approx1.5$, the code with SSPRK is $\sim50~\%$ faster than the code with RK4, as noted in Section \ref{s2.3}.

\begin{figure*}
\centering
\vskip 0.1 cm
\includegraphics[width=0.9\linewidth]{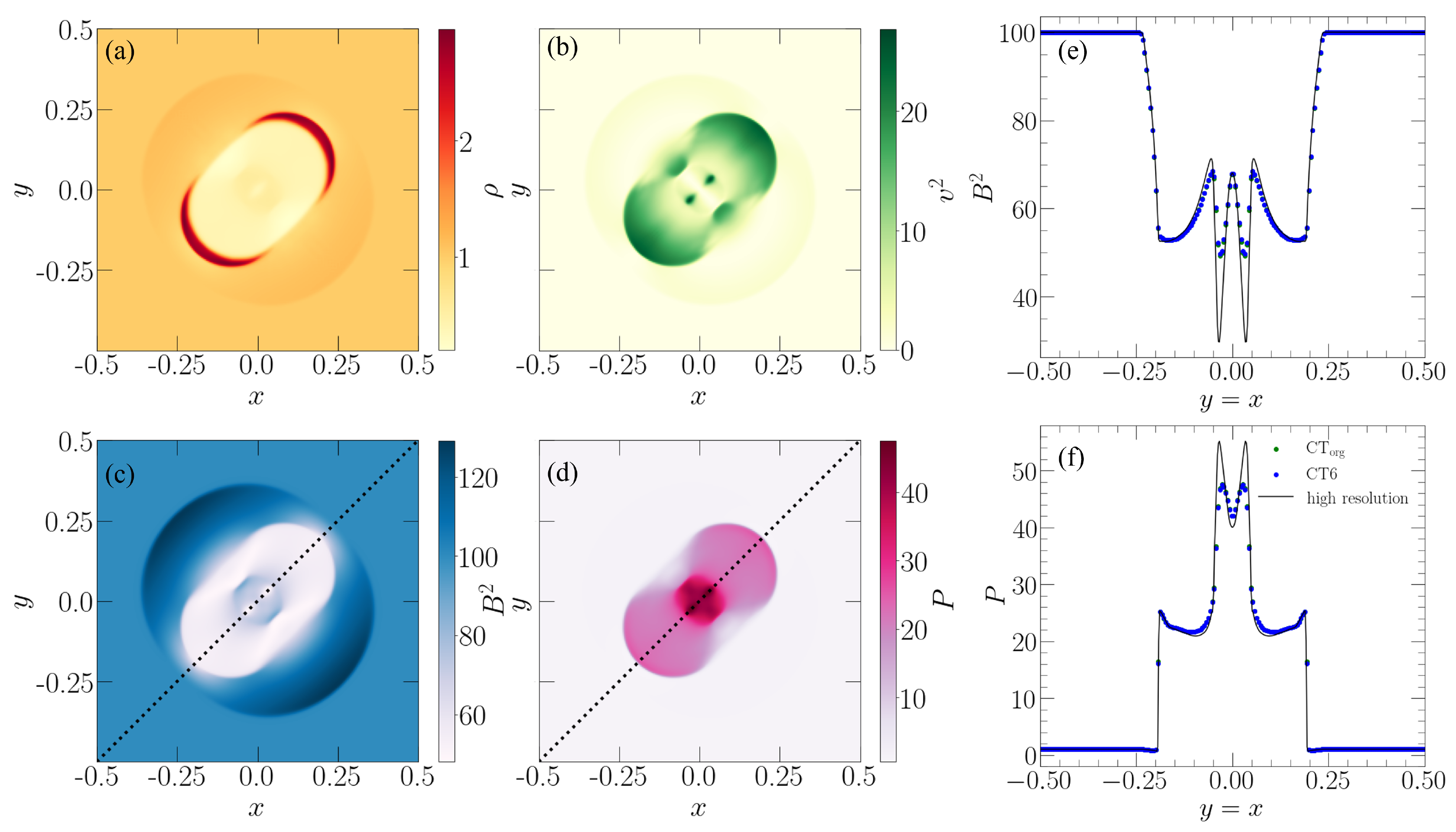}
\vskip -0.1 cm
\caption{Results of the 3D MHD blast wave test. See the main text for the setup. The computational domain consists of $[-0.5,0.5]\times[-0.5,0.5]\times[-0.5,0.5]$ with $200\times200\times200$ grid cells. Panels (a) to (d): The 2D slice images of $\rho$, $v^2$, $B^2$, and $P$ through $z=0$. The results of a simulation at $t=0.02$ using the code with CT6 are shown. Panels (e) and (f): The line profiles of $B^2$ and $P$ along the diagonal line of $x=y$ at $z=0$ (dotted lines in panels (c) and (d)). The results of simulations using the codes with CT6 and CT$_{\rm org}$ are shown. The solid lines draw the results of a higher resolution simulation with $400\times400\times400$ grid cells using the code with CT6 for comparison.}
\label{f12}
\end{figure*}

\subsection{2D MHD Rotor}
\label{s3.6}

We next present the MHD rotor test. It was used to assess the performance of MHD codes in capturing rotational flows with magnetic fields in a number of papers \citep[e.g.,][]{balsara1999,stone2008,toth2000,donnert2019}. Initially, a rotating disk is placed in the 2D computational domain with the magnetic field perpendicular to the rotation axis. The shear at the border of the rotating disk generates rotational discontinuities, and the unbalanced centrifugal force drives the expansion of the disk, producing shocks and rarefaction waves.

Specifically, we consider the first rotor problem \citep[see][]{toth2000}. The initial setup consists of a disk with a radius of 0.1 and a density of $\rho=10$, located in the center of the $x-y$ plane of $[-0.5,0.5]\times[-0.5,0.5]$ with $\rho=1$. The disk rotates with a uniform angular velocity of 20. The pressure, $P=1$, is uniform everywhere, with the adiabatic index $\gamma = 1.4$. A taper of width 0.015 is applied around the disk to smooth the initial discontinuity. The magnetic field is uniform with $\mbox{\boldmath$B$}=B_0\hat{\mbox{\boldmath$x$}}=5/\sqrt{4\pi}\hat{\mbox{\boldmath$x$}}$. Refer \citet{toth2000} for the further details of the initial setup.

In the panels (a) to (d) of Figure \ref{f10}, the distributions of $\rho$, $v^2$, $B^2$, and $P$ from a simulation with $400\times400$ grid cells at $t=0.15$ using the code with CT6 are shown. In panels (e) and (f), $B_y$ along the $x$-axis and $B_x$ along the $y$-axis from simulations with CT6 (blue dots) and CT$_{\rm org}$ (green dots) are compared to those from a higher resolution simulation. The results of the two simulations are basically very similar, with the blue and green dots almost overlapping. The results with CT4, which are not shown, are also basically identical; in the test, the CT algorithm does not make any noticeable differences. All the structures, including shocks, discontinuities, and waves, are well reproduced; almost perfect symmetry is maintained, and no spurious oscillations are observed.

With the uniform magnetic field, initially its divergence is perfectly zero. Hence, the performance of the code to preserve the divergence-free requirement can be examined in this test. Figure \ref{f11} shows the maximum values of the normalized $\mbox{\boldmath$\nabla$}\cdot\mbox{\boldmath$b$}$ in simulations with CT6, CT4, and CT$_{\rm org}$. They are calculated with Equation (\ref{divbFD}) using Equation (\ref{FD}) for the cases of CT6 and CT4 and using a second-order FD for CT$_{\rm org}$. The values increase to nonzero but saturate at the level of numerical truncation error, regardless of the CT algorithms used.  This confirms that all the CT algorithms successfully maintain the divergence-free constraint.

\begin{figure*}
\centering
\vskip 0.1 cm
\includegraphics[width=0.8\linewidth]{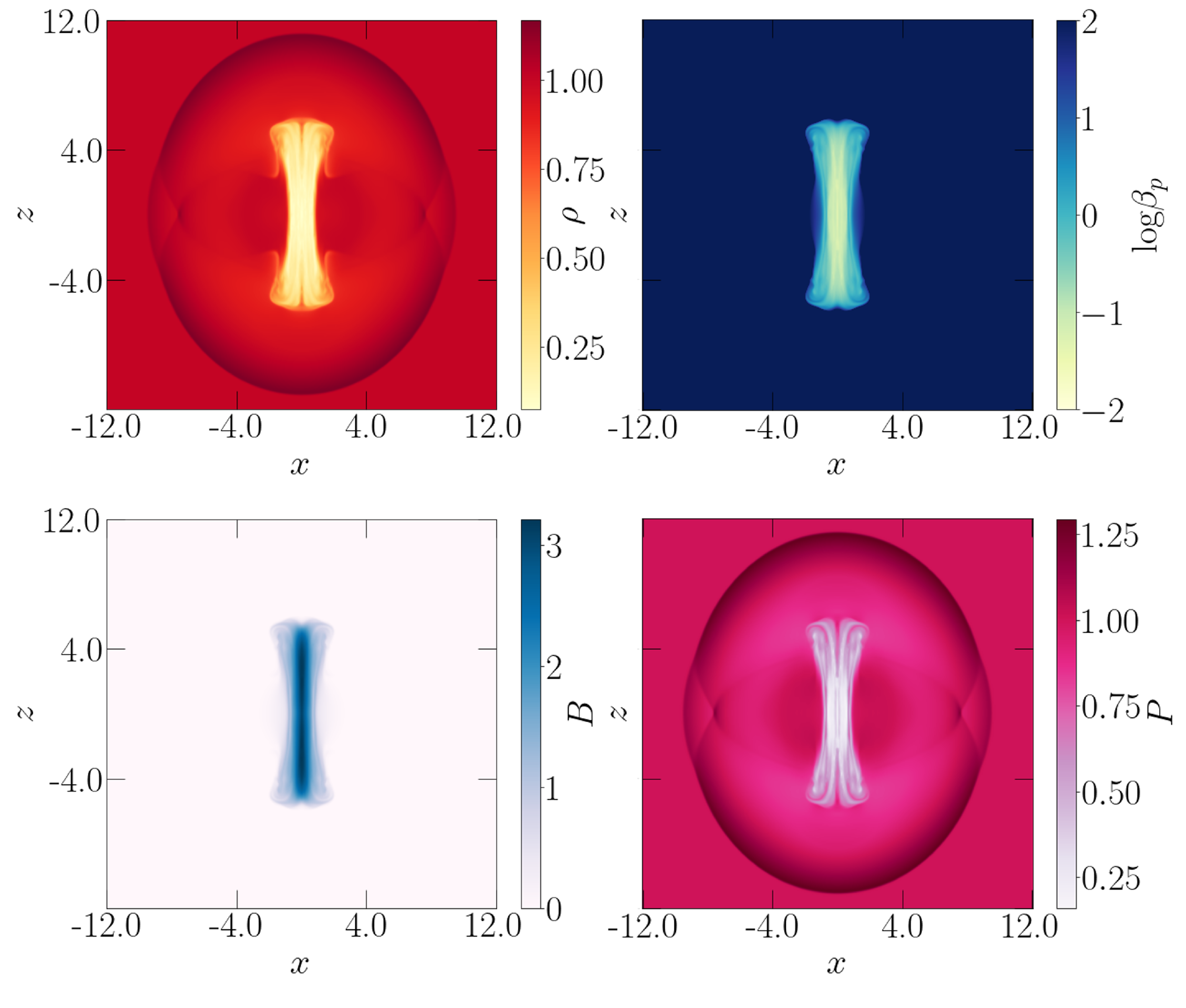}
\vskip -0.1 cm
\caption{Results of the 3D MHD jet launching test. See the main text for the setup. The 2D slice images of $\rho$, $\log(\beta_p)$, $B$, and $P$ through $y=0$ from a simulation using the code with CT6 are shown. The computational domain consists of $[-12,12]\times[-12,12]\times[-12,12]$ with $480\times480\times480$ grid cells, and the images are at the end of the simulation, $t=5$. }
\label{f13}
\end{figure*}

\subsection{MHD Blast Wave in 3D}
\label{s3.7}

MHD blast wave tests can be set up to produce strong shocks and discontinuities in multi-dimensions, and hence they have been used to verify the reliability and robustness of the code. There have been different versions in the literature, and we adopt the blast condition used, for instance, in \citet{londrillo2000}, \citet{stone2008}, and \citet{donnert2019}. In particular, \citet{donnert2019} performed the test in 3D, and we present the same 3D test. The initial condition is given in the computational domain of $[-0.5,0.5]\times[-0.5,0.5]\times[-0.5,0.5]$ as
\begin{equation}
\mbox{\boldmath$u$}=(1,0,0,0,B_0/\sqrt{2},B_0/\sqrt{2},0,P)^T, 
\end{equation}
with the initially uniform magnetic field of $B_0=10$ along the diagonal direction in the $x-y$ plane. The blast wave is generated by the high pressure in the center of the computational domain:
\begin{equation}
P=
\begin{cases}
1 & r\le r_0 \\
1+99f(r) & r_0<r\le r_1, \\
100 & r>r_1
\end{cases}
\end{equation}
where $r=\sqrt{x^2+y^2+z^2}$, $r_0=0.125$, $r_1=1.1~r_0$, and $f=(r_1-r)/(r_1-r_0)$. As in the MHD rotor test, a taper is applied around the press jump for a smooth start-up.

Figure \ref{f12} shows the results of simulations with $200\times200\times200$ grid zones at $t=0.02$. In panels (a) to (d), the 2D distributions of $\rho$, $v^2$, $B^2$, and $P$ through $z=0$ from a simulation using the code with CT6 are drawn. In panels (e) and (f), the line profiles of $B^2$ and $P$ along $x=y$ at $z=0$ from simulations with CT6 (blue dots) and CT$_{\rm org}$ (green dots) are plotted and compared to those of a higher resolution simulation with $400\times400\times400$ grid zones.  The axisymmetry around the $x=y$ diagonal line is well preserved in 3D, and no spurious oscillations are observed. All the structures, including strong shocks, are well captured, whereas local peaks are somewhat underrepresented in simulations with $200\times200\times200$ grid zones.

\begin{figure*}
\centering
\vskip 0.1 cm
\includegraphics[width=0.75\linewidth]{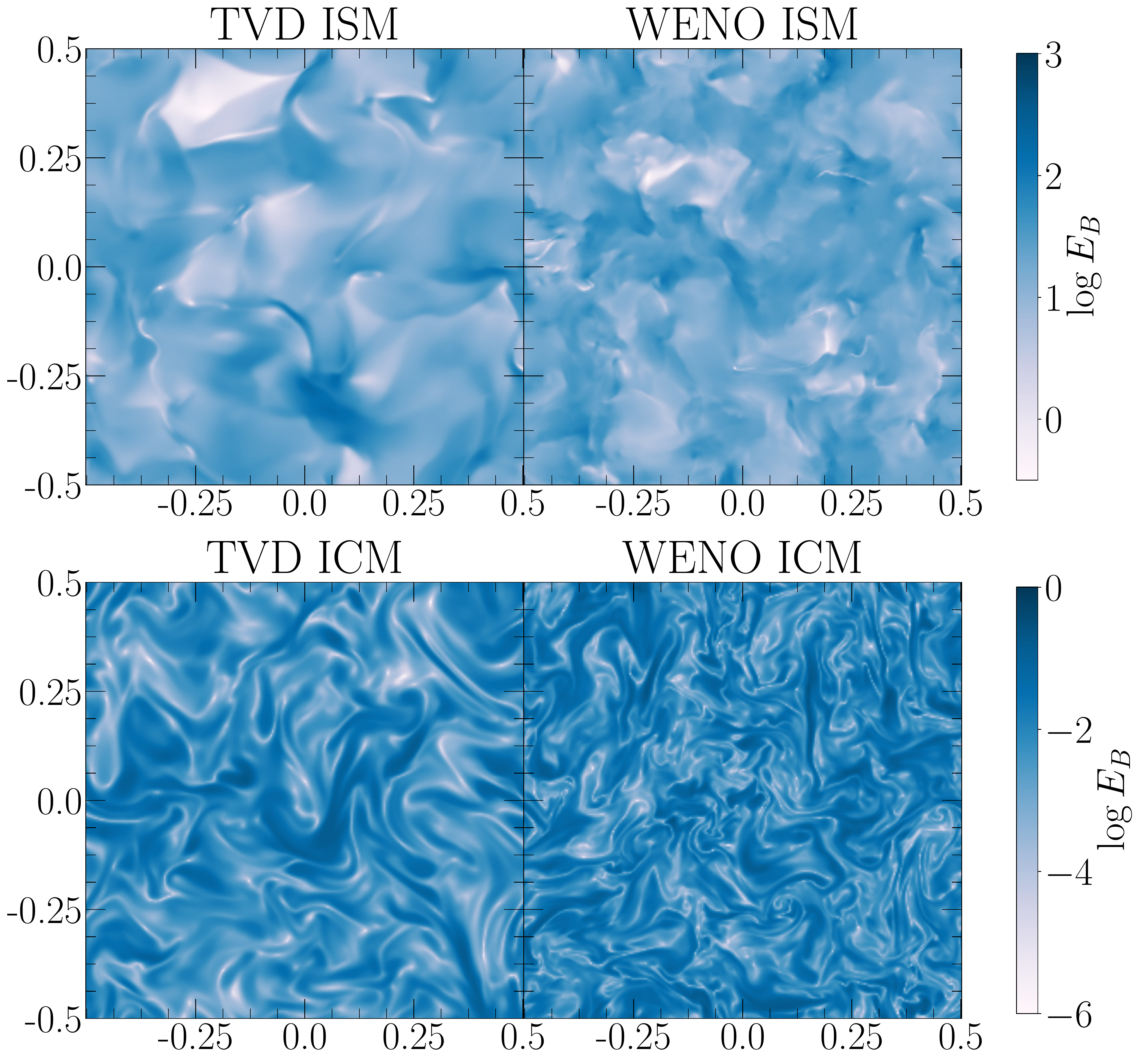}
\vskip -0.1 cm
\caption{2D slice images of the magnetic energy in the ISM turbulence with $M_{\rm turb}\approx 10$ and $\beta_p=0.1$ (top panels) and the ICM turbulence with $M_{\rm turb}\approx 0.5$ and $\beta_p=10^6$ (bottom panels), from simulations with $256\times256\times256$ grid cells using the TVD code (left panels) and the WENO code with CT6 (right panels), at the end of simulations. See the main text for the setup of the turbulences.}
\label{f14}
\end{figure*}

\begin{figure*}
\vskip 0.1 cm
\hskip -0.7 cm
\includegraphics[width=1.05\linewidth]{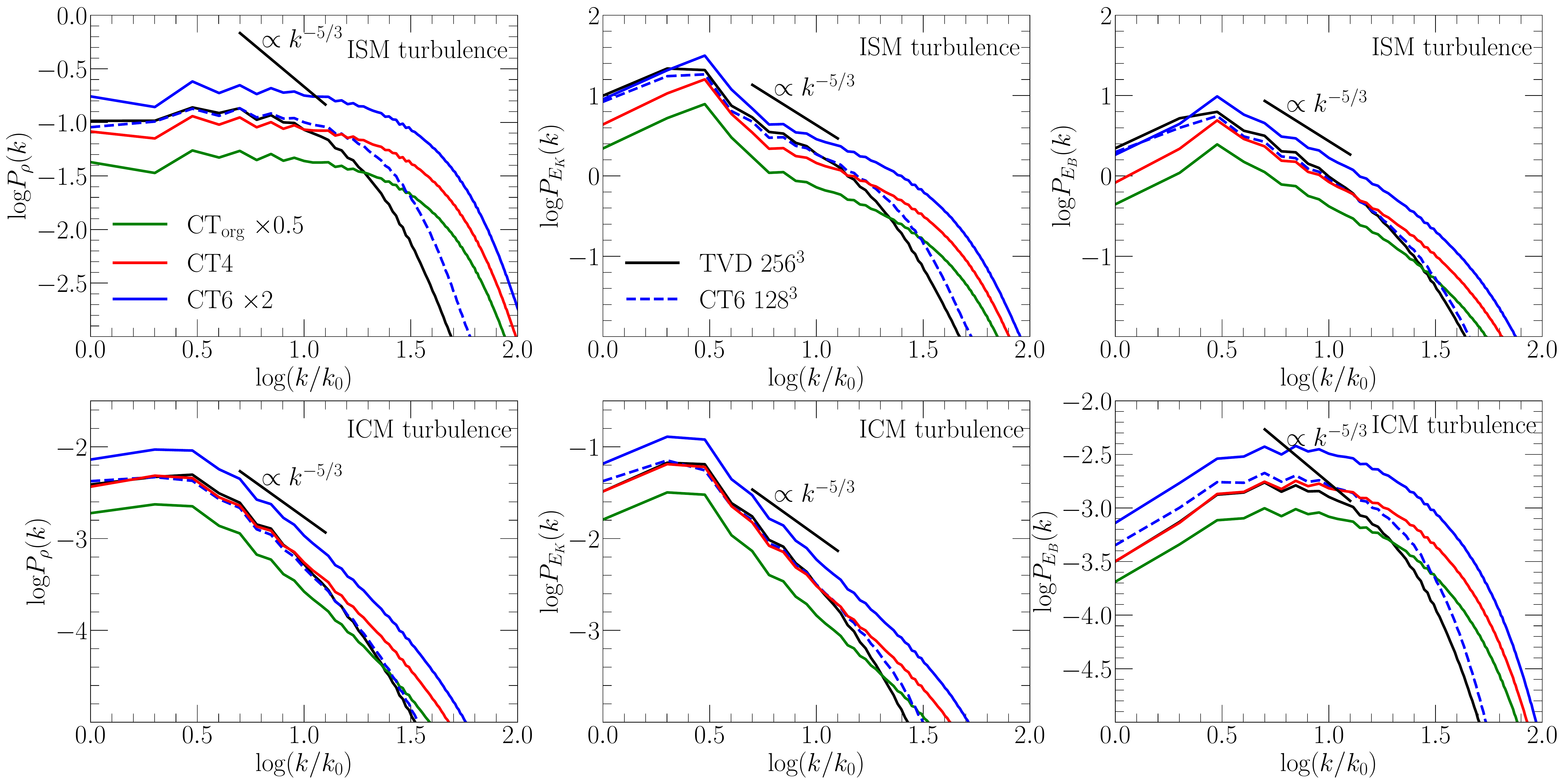}
\vskip -0.1 cm
\caption{Power spectra of the density, $P_{\rho}(k)$ (left panels), the kinetic energy, $P_{E_K}(k)$ (middle panels), and the magnetic energy, $P_{E_B}(k)$ (right panels), in the ISM turbulence with $M_{\rm turb}\approx 10$ and $\beta_p=0.1$ (top panels) and the ICM turbulence with $M_{\rm turb}\approx 0.5$ and $\beta_p=10^6$ (bottom panels). See the main text for the setup of the turbulences. The results from simulations with $256\times256\times256$ grid cells using the WENO codes with CT6, CT4, and CT$_{\rm org}$ and also using the TVD code are shown. The spectra are averaged over the saturated period. To avoid overlap, the spectra with the codes with CT6 and CT$_{\rm org}$ are multiplied by factors of 2 and 1/2, respectively. For comparison purposes, the results from a lower resolution simulation with $128\times128\times128$ grid cells using the WENO code with CT6 are shown with dashed lines. The short black lines draw the slope of the Kolmogorov spectrum.}
\label{f15}
\end{figure*}

\subsection{3D MHD Jet Launching}
\label{s3.8}

Considering that jets are common in astrophysical environments, we add a test to follow the launch and early evolution of MHD jets in 3D. Specifically, we consider the impulsive jets suggested by \citet{li2006}, which model the jets produced by strong magnetic fields generated at a black hole. The initial state is given as
\begin{equation}
\mbox{\boldmath$u$}=(1,0,0,0,B_x,B_y,B_z,1)^T.
\end{equation}
The magnetic field is set up with the vector potential
\begin{equation}
\begin{aligned}
A_x = -\exp(-r^2) y ~~ \\
A_y = \exp(-r^2) x ~~~~ \\
A_z = 0.5A_0\exp(-r^2),
\end{aligned}
\end{equation}
at grid cell edges; $\mbox{\boldmath$b$}$ is calculated with $\mbox{\boldmath$A$}$ at grid cell interfaces, and then $\mbox{\boldmath$B$}$ is calculated at grid centers. Here, $r=\sqrt{x^2+y^2+z^2}$, and $A_0=20$ is used. The computational domain consists of $[-12,12]\times[-12,12]\times[-12,12]$ with $480\times480\times480$ grid cells, and has outflow boundaries. The initial field does not satisfy the force equilibrium because of the presence of the nonzero Lorentz force, $\mbox{\boldmath$J$}\times \mbox{\boldmath$B$}$. The Lorentz force drives the convergence of the fluid, inducing the launch of a pair of light magnetized jets along the $z$-axis.

Figure \ref{f13} depicts the 2D distributions of $\rho$, $\log(\beta_p)$, $B$, and $P$ through $y=0$ at $t=5$ from a simulation using the code with CT6. All the structures, including the elongated jets of helical magnetic fields, are well reproduced, and the axisymmetry around the $z$-axis is nearly preserved in 3D. On the other hand, there are structures around and inside the jets, some of which are possibly the consequences of instabilities \citep[see][for further discussion]{li2006}.

\begin{figure*}
\centering
\vskip 0.1 cm
\includegraphics[width=0.75\linewidth]{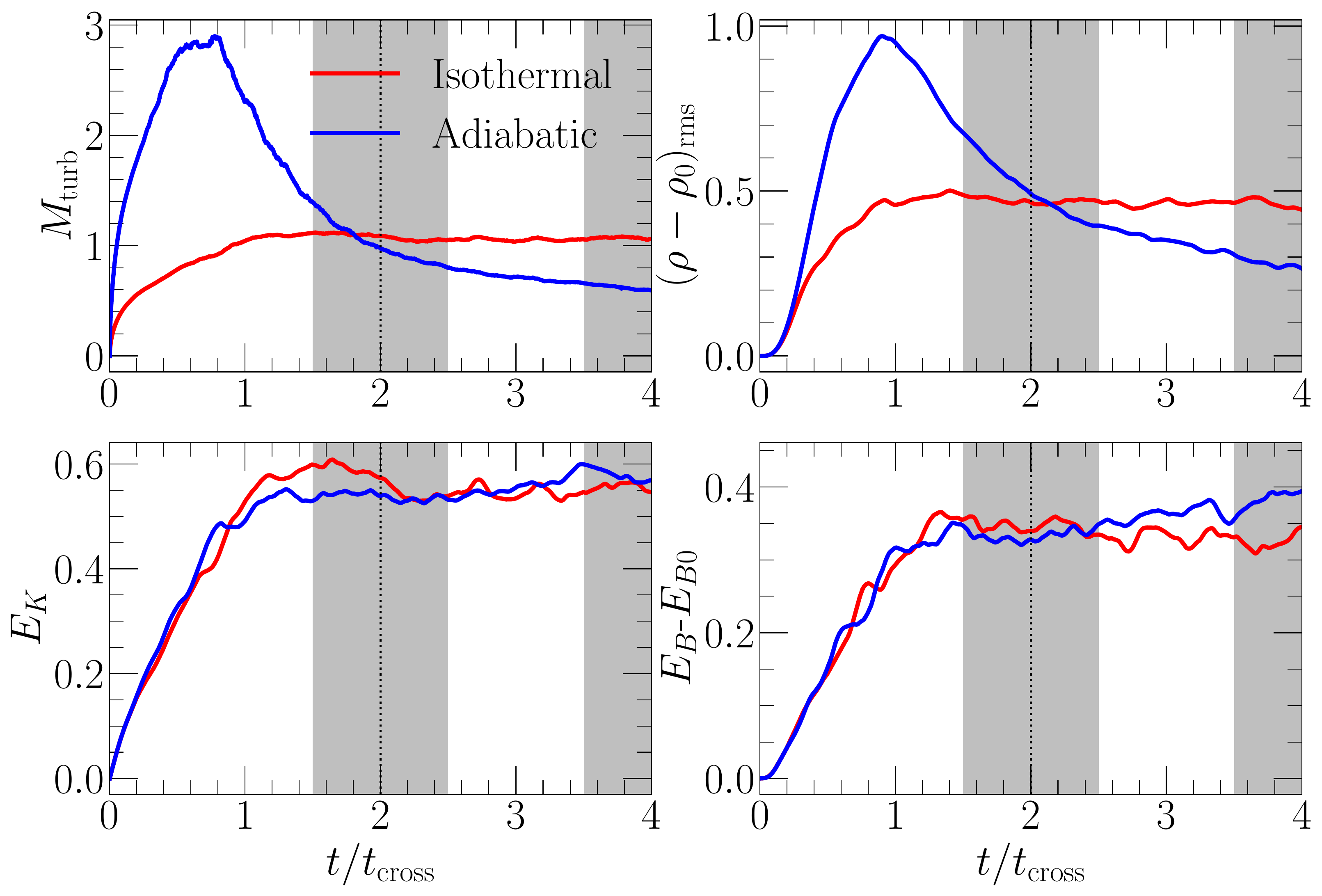}
\vskip -0.1 cm
\caption{Time evolution of $M_{\rm turb}$, $(\rho-\rho_0)_{\rm rms}$, the kinetic energy, and the magnetic energy in the isothermal turbulence with the isothermal MHD code (red lines) and the adiabatic turbulence with the adiabatic MHD code (blue lines). See the main text for the setup of the turbulences. The results from simulations with $256\times256\times256$ grid cells using the isothermal and adiabatic codes with CT6 are shown. The vertical dotted lines mark the time when the images in Figure \ref{f17} are drawn. The shaded regions cover the periods over which the power spectra in Figure \ref{f18} are calculated.}
\label{f16}
\end{figure*}

\begin{figure*}
\centering
\vskip 0.1 cm
\includegraphics[width=0.85\linewidth]{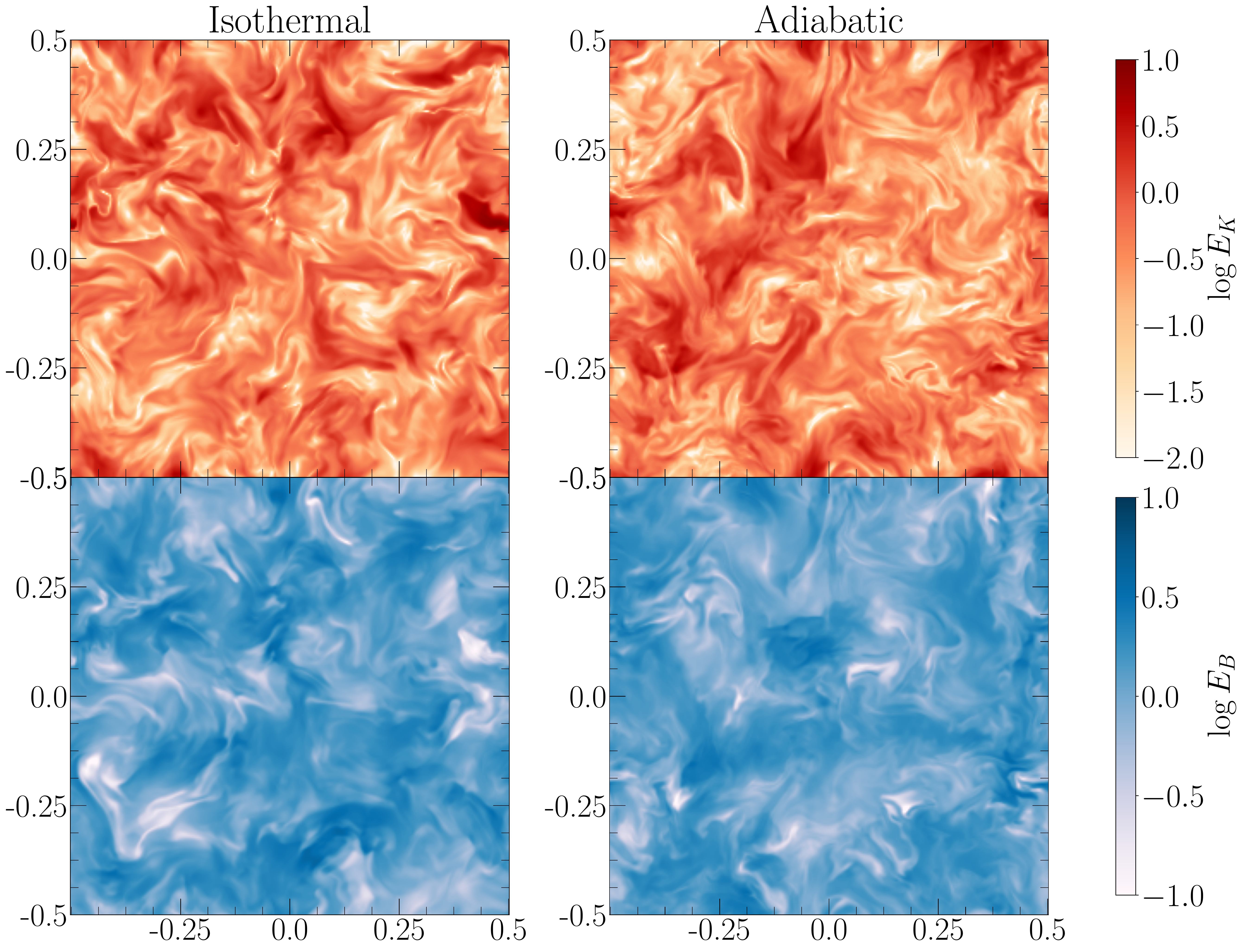}
\vskip -0.1 cm
\caption{2D slice images of the kinetic energy (top panels) and the magnetic energy (bottom panels) in the isothermal turbulence with the isothermal MHD code (left panels) and the adiabatic turbulence with the adiabatic MHD code (right panels), both having $M_{\rm turb}\approx 1$ and $\beta_p=1$, from simulations with $256\times256\times256$ grid cells using codes with CT6. See the main text for the setup of the turbulences.}
\label{f17}
\end{figure*}

\begin{figure*}
\vskip 0.1 cm
\hskip -0.7 cm
\includegraphics[width=1.05\linewidth]{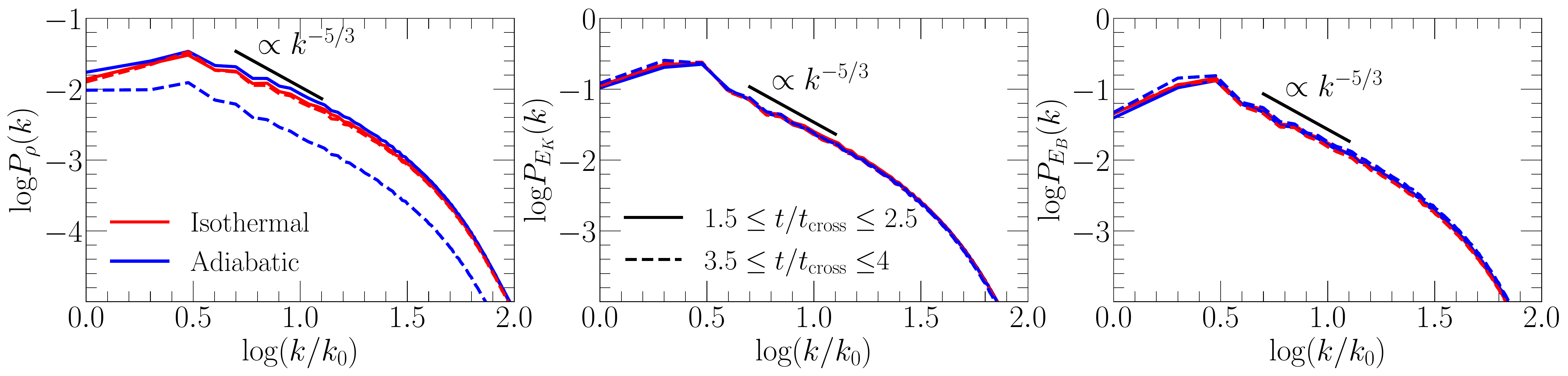}
\vskip -0.1 cm
\caption{Power spectra of the density, $P_{\rho}(k)$ (left panel), the kinetic energy, $P_{E_K}(k)$ (middle panel), and the magnetic energy, $P_{E_B}(k)$ (right panel), in the isothermal turbulence with the isothermal MHD code (red lines) and the adiabatic turbulence with the adiabatic MHD code (blue lines). See the main text for the setup of the turbulences. The results from simulations with $256\times256\times256$ grid cells using the isothermal and adiabatic codes with CT6 are shown. The solid lines show the spectra averaged over $1.5\le t/t_{\rm cross}\le 2.5$, during which both the turbulences are characterized by $M_{\rm turb}\approx 1$ and $\beta_p=1$. The dashed lines draw the spectra averaged over $3.5\le t/t_{\rm cross}\le 4$, when $M_{\rm turb}\approx 1$ for the isothermal turbulence but $M_{\rm turb} < 1$ for the adiabatic turbulence. The short black lines draw the slope of the Kolmogorov spectrum.}
\label{f18}
\end{figure*}

\subsection{MHD Turbulence 1}
\label{s3.9}

In astrophysics, isothermal MHD codes have been widely applied to studies of turbulence, as noted in the introduction. Hence, as a test of our isothermal MHD code, we present simulations of turbulent flows. We also compare the results to those with the isothermal MHD code based on the TVD scheme of second-order accuracy (referred as the TVD code) \citep{kim1999}, which was previously used for studies of astrophysical turbulence \citep[e.g,][]{porter2015,park2019}.

We consider two cases that are identified by the following parameters of turbulence: $M_{\rm turb}\approx 10$ and $\beta_p=0.1$ and $M_{\rm turb}\approx 0.5$ and $\beta_p=10^6$. Here, $M_{\rm turb}\equiv v_{\rm rms}/a$ with the root-mean-square (rms) fluid velocity $v_{\rm rms}$ and the sound speed $a$ is the turbulent Mach number, and $\beta_p$ is the initial plasma beta. The first case intends to model the turbulence in ISM molecular clouds (referred as the ISM turbulence), whereas the second case intends to model the turbulence in the ICM (referred as the ICM turbulence).

The initial state is given as
\begin{equation}
\mbox{\boldmath$u$}=(1,0,0,0,B_0,0,1)^T, \label{turninitst}
\end{equation}
where $B_0$ is assigned according to $\beta_p$. Turbulence is driven with solenoidal forcing ($\mbox{\boldmath$\nabla$}\cdot\delta\mbox{\boldmath$v$}=0$). Velocity perturbations, $\delta\mbox{\boldmath$v$}$, are drawn from a Gaussian random field with the spectrum of 
\begin{equation}
|\delta\mbox{\boldmath$v$}_k|^2\propto k^6 \exp\left(-\frac{8k}{k_{\rm exp}}\right)~~{\rm with}~~k_{\rm exp}=
\frac{4\pi}{L_0}, \label{drvposp}
\end{equation}
where $L_0$ is the size of computational box. The amplitude of $\left<|\delta\mbox{\boldmath$v$}|\right>$ is constant in time, and the driving is temporally uncorrelated with the perturbations drawn randomly at each time step. See \citet{park2019} and \citet{cho2022} for the further details of turbulence driving. Simulations are run in the periodic computational domain of $[-0.5,0.5]\times[-0.5,0.5]\times[-0.5,0.5]$ with $256\times256\times256$ grid cells, up to $t=5~t_{\rm cross}$ for the ISM turbulence and $t=30~t_{\rm cross}$ for the ICM turbulence, using our code based on the WENO scheme as well as the TVD code. Here, $t_{\rm cross}$ is the crossing time defined as $L_{\rm inj}/v_{\rm rms}$, and $L_{\rm inj}$ is the injection scale roughly given as $L_0/2$ in Equation (\ref{drvposp}).

Figure \ref{f14} depicts the 2D slice distributions of the magnetic energy ($E_B=B^2/2$) in the ISM turbulence (top panels) and the ICM turbulence (bottom panels), from simulations using the TVD code (left panels) and the WENO code with CT6 (right panels), at the end of simulations. The images reveal the characteristics of the turbulences; in the ISM turbulence with high $M_{\rm turb}$, shocks are apparent \citep[see, e.g.,][]{cho2022}, whereas in the ICM turbulence with $M_{\rm turb}<1$, flux tubes or cross sections of flux ribbons are visible \citep[see, e.g.,][]{porter2015}. A noticeable difference between the results with the WENO and TVD codes is the presence of small scale structures; there are more small scale structures with the WENO code, reflecting the higher-order nature of the code, than with the TVD code.

To further examine the impact of the accuracy order of the codes, the power spectra of the density, kinetic energy ($E_K=\rho v^2/2$), and magnetic energy, averaged during the saturated stage, are plotted in Figure \ref{f15}; the top panels present the power spectra of the ISM turbulence over the time interval of $2.5~t_{\rm cross}\le t \le 5~t_{\rm cross}$, and the bottom panels present the power spectra of the ICM turbulence over $15~t_{\rm cross}\le t \le 30~t_{\rm cross}$. The results from simulations using the codes with CT6, CT4, and CT$_{\rm org}$ as well as the TVD code are shown; the results of a lower resolution simulation with $128\times128\times128$ grid cells using the code with CT6 are also shown for comparison. The spectra are typical of the ISM and ICM turbulences and are consistent with those of previous works \cite[see, e.g.,][]{porter2015,cho2022}. The spectra with CT6, CT4, and CT$_{\rm org}$ are basically identical, suggesting that turbulence simulations are not sensitive to the CT algorithm as long as the upwind scheme for the calculation of numerical fluxes is the same. On the other hand, with the WENO code, the spectra extend to higher wavenumbers than with the TVD code. As a matter of fact, the spectra with the TVD code look comparable to those of the lower resolution simulation with the WENO code. This indicates that the WENO code has a higher ``effective resolution'' owing to its higher-order accuracy, and hence potentially can have a higher computational efficiency, than the TVD code.

\subsection{MHD Turbulence 2}
\label{s3.10}

Although isothermal MHD codes have been primarily used for previous studies of astrophysical turbulence, the isothermal approximation holds only when cooling is highly efficient, as mentioned in the introduction. The effects of adiabaticity, on the other hand, have been investigated using adiabatic codes in studies such as \citet{nolan2015} for hydrodynamic turbulent flows and \citet{grete2020} for MHD turbulent flows.

We here present test simulations of turbulent flows using both our isothermal and adiabatic MHD codes, comparing the performance of the two codes as well as the properties of isothermal and adiabatic turbulences. We intend to produce isothermal and adiabatic turbulences of $M_{\rm turb}\approx 1$ and $\beta_p=1$ for this test. The initial state is again given in Equation (\ref{turninitst}), and turbulence is driven in the same way as in the previous section. The simulation box is also the same as in the previous section, and simulations are run using isothermal and adiabatic MHD codes with CT6, up to $t=4~t_{\rm cross}$. Cooling is not included in the simulation of adiabatic turbulent flows.

Figure \ref{f16} shows the time evolution of flow quantities, $M_{\rm turb}$, $(\rho-\rho_0)_{\rm rms}$, $E_K$, and $E_B-E_{B0}$, where the subscript $0$ denotes the initial values. The isothermal turbulence saturates, reaching a statistically steady state, at $t\gtrsim1.5~t_{\rm cross}$. In contrast, the adiabatic turbulence, without cooling included, does not exhibit saturation; $M_{\rm turb}$ and $(\rho-\rho_0)_{\rm rms}$, after the initial growth, continue to decrease over time. Nevertheless, at the time when $M_{\rm turb}$ is around unity in both the isothermal and adiabatic turbulences, that is, at $1.5\lesssim t/t_{\rm cross}\lesssim2.5$, the flow quantities, including $(\rho-\rho_0)_{\rm rms}$, are comparable in the two turbulences.

Figure \ref{f17} illustrates the 2D slice distributions of $E_K$ (top panels) and $E_B$ (bottom panels) in the isothermal turbulence (left panels) and the adiabatic turbulence (right panels) at $t=2~t_{\rm cross}$. With similar $M_{\rm turb}$, the two types of turbulence look very similar.

In Figure \ref{f18}, the power spectra of the density, kinetic energy, and magnetic energy, averaged over $1.5\lesssim t/t_{\rm cross}\lesssim2.5$, are plotted with solid lines. The power spectra of the two turbulences almost overlap, and the similarity between the two turbulences is confirmed. In Figure \ref{f18}, the power spectra at $3.5\lesssim t/t_{\rm cross}\lesssim4$ are also shown with dashed lines. With smaller $(\rho-\rho_0)_{\rm rms}$ in Figure \ref{f17}, the amplitude of the density power in the adiabatic turbulence is smaller than in the isothermal turbulence, but the shape, including the peak and slope, is still similar in the two turbulences. With similar $E_K$ and $E_B$, the energy power spectra at $3.5\lesssim t/t_{\rm cross}\lesssim4$ are similar in the two turbulences. Our results are consistent with those of \citet{grete2020} who claimed that the properties of turbulence are not sensitive to the EoS in the MHD equations. Our results also indicate that both our isothermal and adiabatic MHD codes have similar accuracy in simulations of turbulent flows.

\section{Summary}
\label{s4}

We have introduced HOW-MHD, a new MHD code for astrophysical applications. It is based on the FD (finite difference) WENO scheme of fifth-order spatial accuracy and the five-stage SSPRK time-integration method of fourth-order temporal accuracy. Most of all, the code is equipped with a newly developed CT (constrained transport) algorithm of high-order accuracy for the divergence-free constraint of magnetic fields. \textcolor{black}{The algorithm utilizes a high-order FD method, and is named according to the order of the differencing, CT4 and CT6.} All together, the code achieves high-order accuracy, as well as high reliability and robustness. Both the adiabatic and isothermal versions of the MHD code have been described.

1. In \textcolor{black}{the} tests involving the propagation of linear MHD waves and a circularly polarized Alfv\'en wave in 3D, the code achieves the fifth-order convergence in the damping of the waves. In \textcolor{black}{the} tests involving the advection of a magnetic loop and the reconnection of magnetic fields, we observe that the numerical diffusivity  of the code effectively decreases \textcolor{black}{when CT6 and CT4 are employed}.

2. In an oblique shock-tube test, the ability of the code to accurately capture shocks and discontinuities in multi-dimensions is verified. In addition, it is shown that with SSPRK, which enables CFL $\geq1$, the computational efficiency increases, without degrading the quality of capturing shocks and discontinuities. In test simulations of MHD rotor, MHD blast wave, and MHD jet launching, the ability of the code to handle strong shocks and complex flows is proved.

3. With \textcolor{black}{the} tests involving turbulent flows, it is shown that our isothermal MHD code has high effective resolutions owing to its high-order accuracy, potentially improving the computational efficiency. In addition, by comparing the isothermal turbulence using the isothermal MHD code and the adiabatic turbulence using the adiabatic MHD code, it is shown that the two codes have similar accuracy.

Overall, with its high-order accuracy, the new MHD code, HOW-MHD, should have the potential to be a valuable tool for studying the complex processes that govern the behavior of magnetized fluids in the universe.

Finally, we note that the code presented here is for Cartesian geometry. In principle, it is possible to extend the code to other geometries, such as cylindrical and spherical geometries, by properly implementing the volume effects. However, this will require careful approaches to maintain high-order accuracy if the code is high-order. We leave the development of the cylindrical and spherical versions of our MHD code for future work.

\begin{acknowledgments}

This work was supported by the National Research Foundation (NRF) of Korea through grant 2020R1A2C2102800. Some of simulations were performed using the high performance computing resources of the UNIST Supercomputing Center.

\end{acknowledgments}

\newpage

\appendix

\section{MHD Equations and Eigenvalues}
\label{sa1}

\subsection{Adiabatic MHDs}
\label{sa1.1}

The ideal magnetohydrodynamic (MHD) equations for adiabatic flows are given as
\begin{eqnarray}
\frac{\partial \rho}{\partial t} + \mbox{\boldmath$\nabla$}\cdot(\rho\mbox{\boldmath$v$}) = 0,~~~~~~~~~~~~~~~~\\ \label{mass}
\frac{\partial \mbox{\boldmath$v$}}{\partial t} + \mbox{\boldmath$v$}\cdot\mbox{\boldmath$\nabla$}\mbox{\boldmath$v$}+\frac{1}{\rho}\mbox{\boldmath$\nabla$}P -\frac{1}{\rho}(\mbox{\boldmath$\nabla$}\times\mbox{\boldmath$B$})\times\mbox{\boldmath$B$}=0,\\ \label{acceleration}
\frac{\partial B}{\partial t} - \mbox{\boldmath$\nabla$}\times(\mbox{\boldmath$v$}\times\mbox{\boldmath$B$}) = 0,~~~~~~~~~~~~\\ \label{induction}
\frac{\partial P}{\partial t} + \mbox{\boldmath$v$}\cdot\mbox{\boldmath$\nabla$}P+\gamma P\mbox{\boldmath$\nabla$}\cdot\mbox{\boldmath$v$}=0,~~~~~~~~~\label{pressure}
\end{eqnarray}
where $\rho$, \mbox{\boldmath$v$}, \mbox{\boldmath$B$}, $P$, and $\gamma$ are the density, velocity, magnetic field, pressure, and adiabatic index, respectively. Here, the units are chosen so that $4\pi$ does not appear in the equations.

They are written in conservative form as
\begin{eqnarray}
\frac{\partial \mbox{\boldmath$q$}}{\partial t}+\frac{\partial \mbox{\boldmath$F$}}{\partial x}+\frac{\partial \mbox{\boldmath$G$}}{\partial y}+\frac{\partial \mbox{\boldmath$H$}}{\partial z}=0, \label{3dFD}
\end{eqnarray}
in Cartesian geometry. Here, \mbox{\boldmath$q$} is the state vector of conserved quantities, given as,
\begin{eqnarray}
\mbox{\boldmath$q$}=(\rho,\rho v_x,\rho v_y,\rho v_z,B_x,B_y,B_z,E)^T,
\end{eqnarray}
and \mbox{\boldmath$F$}, \mbox{\boldmath$G$}, and \mbox{\boldmath$H$} are the flux vectors in the $x$-, $y$-, and $z$-directions, respectively. For instance, \mbox{\boldmath$F$} is given as
\begin{equation}
\mbox{\boldmath$F$}=\begin{pmatrix} \rho v_x  \\ \rho v_x^2+P^*-B_x^2 \\ \rho v_x v_y - B_xB_y \\ \rho v_x v_z - B_xB_z \\ 0 \\ B_yv_x-B_xv_y \\ B_zv_x-B_xv_z \\ 
(E+P^*)v_x-B_x(B_xv_x+B_yv_y+B_zv_z) \end{pmatrix},\label{MHDflux1}
\end{equation}
and \mbox{\boldmath$G$} and \mbox{\boldmath$H$} are given by properly permuting indices. The total pressure $P^*$ and the total energy $E$ are given as
\begin{eqnarray}
P^* & = & P+\frac{1}{2}B^2,\\
E & = & \frac{P}{\gamma-1}+\frac{1}{2}(\rho v^2+B^2),
\end{eqnarray}
where $B^2=B_x^2+B_y^2+B_z^2$ and $v^2=v_x^2+v_y^2+v_z^2$.

Along each direction, there are seven characteristic modes in adiabatic MHDs. Their eigenvalues along the $x$-direction, $\lambda_x^1, \cdots, \lambda_x^7$ in non-increasing order, are
\begin{eqnarray}
\lambda_x^{1,7} & = & v_x \pm c_f,\\
\lambda_x^{2,6} & = & v_x \pm c_A,\\
\lambda_x^{3,5} & = & v_x \pm c_s,\\
\lambda_x^4 & = & v_x.
\end{eqnarray}
Here, $c_f$, $c_A$, and $c_s$ are the speeds of the fast, Alfv\'en, and slow waves, respectively, which are given as
\begin{eqnarray}
c_{f,s} = \left\{\frac{1}{2}\left[a^2+\frac{B^2}{\rho}\pm\sqrt{\left(a^2+\frac{B^2}{\rho}\right)^2-4a^2\frac{B_x^2}{\rho}}\right]\right\}^{1/2}~~~\\
c_A = \left(\frac{B_x^2}{\rho}\right)^{1/2},~~~~~~~~~~~~~~~~~~~~~~~~~~
\end{eqnarray}
where $a$ is the sound speed
\begin{equation}
a = \left(\gamma\frac{P}{\rho}\right)^{1/2}.
\end{equation}
The eigenvalues along the $y$- and $z$-directions can be obtained by properly permuting indices.

The left and right eigenvectors are given in the literature, for instance, in \citet{ryu1995a}.

\subsection{Isothermal MHDs}
\label{sa1.2}

The ideal magnetohydrodynamic (MHD) equations for isothermal flows are written as
\begin{eqnarray}
\frac{\partial \rho}{\partial t} + \mbox{\boldmath$\nabla$}\cdot(\rho\mbox{\boldmath$v$}) = 0,~~~~~~~~~~~~~~~~~~ \label{mass2} \\
\frac{\partial \mbox{\boldmath$v$}}{\partial t} + \mbox{\boldmath$v$}\cdot\mbox{\boldmath$\nabla$}\mbox{\boldmath$v$}+\frac{1}{\rho}\mbox{\boldmath$\nabla$}P
-\frac{1}{\rho}(\mbox{\boldmath$\nabla$}\times\mbox{\boldmath$B$})\times\mbox{\boldmath$B$}=0,~~~~ \label{acceleration2}\\
\frac{\partial B}{\partial t} - \mbox{\boldmath$\nabla$}\times(\mbox{\boldmath$v$}\times\mbox{\boldmath$B$}) = 0,~~~~~~~~~~~~~~ \label{MHDflux2}
\end{eqnarray}
where the pressure is given as $P=\rho a^2$ with a pre-allocated, constant sound speed $a$. The conservative form can be written the same as for the adiabatic case, except that the state and flux vectors have only the first seven rows.

Along each direction, there are six characteristic modes. Their eigenvalues along the $x$-direction in non-increasing order are
\begin{eqnarray}
\lambda_x^{1,6} & = & v_x \pm c_f,\\
\lambda_x^{2,5} & = & v_x \pm c_A,\\
\lambda_x^{3,4} & = & v_x \pm c_s.\\
\end{eqnarray}
Note that the entropy mode with the characteristic speed $v_x$ is absent. The speeds of the fast, Alfv\'en, and slow waves are the same as in the adiabatic case, except that the sound speed is constant.

For the left and right eigenvectors, we refer to \citet{kim1999}.

\bibliography{MHDcode}{}
\bibliographystyle{aasjournal}

\end{document}